%% file: main.tex
\documentclass[default]{sn-jnl}

\usepackage{amsthm}%
\usepackage{mathrsfs}%
\usepackage{manyfoot}%
\usepackage{booktabs}%
\usepackage{algorithm}%
\usepackage{algorithmicx}%
\usepackage{algpseudocode}%
\usepackage{listings}%

\usepackage{amsmath,amssymb,amsfonts}
\usepackage{textcomp}
\usepackage{caption}
\usepackage{threeparttable}
\usepackage{subcaption}
\usepackage{multirow}
\usepackage{multicol}
\usepackage{graphicx}
\usepackage[table,xcdraw]{xcolor}
\usepackage{xurl}
\usepackage{comment}
\usepackage{xspace}
\usepackage{booktabs}
\usepackage{algorithmicx}
\usepackage{arydshln}

\newcommand{\SEGMCOMP}{\textsc{Segm\_Comp}\xspace}
\newcommand{\SEGMPROF}{\textsc{Segm\_Prof}\xspace}
\newcommand{\SEGMBALANCED}{\textsc{Segm\_Balanced}\xspace}

\begin{document}

\title{Balanced segmentation of CNNs for multi-TPU inference}

\author*{\fnm{Jorge} \sur{Villarrubia}}\email{jorvil01@ucm.es}

\author{\fnm{Luis} \sur{Costero}}\email{lcostero@ucm.es}

\author{\fnm{Francisco D.} \sur{Igual}}\email{figual@ucm.es}

\author{\fnm{Katzalin} \sur{Olcoz}}\email{katzalin@ucm.es}

\affil{\orgdiv{Dept.} \orgname{Arquitectura de Computadores y Autom\'atica, Universidad Complutense de Madrid}, %
\orgaddress{\city{Madrid}, \country{Spain}}}

\abstract{%
In this paper, we propose different alternatives for CNN (Convolutional Neural Networks)
segmentation, addressing inference processes on computing architectures composed by 
multiple Edge TPUs.  Specifically, we compare the inference performance for a number
of state-of-the-art CNN models taking as a reference inference times on one TPU and
a compiler-based pipelined inference implementation as provided by the Google's Edge TPU
compiler. Departing from a profiled-based segmentation strategy, we provide further
refinements to balance the workload across multiple TPUs, leveraging their co-operative
computing power, reducing work imbalance and alleviating the memory access bottleneck 
due to the limited amount of on-chip memory per TPU. The observed performance results
compared with a single TPU yield super-linear speedups and accelerations up to
$2.60\times$ compared with the segmentation offered by the compiler targeting multiple TPUs.
}


\maketitle

\input body

\section*{Declarations}

\subsection*{Ethical Approval}

Not applicable.

\subsection*{Competing interests}

There are no competing interests.

\subsection*{Authors' contributions}

J.V. conducted the design, implementation and evaluation of the segmentation strategies described in the paper, and collaborated in the writing of the manuscript.
L.C. collaborated in the design and critical analysis of the experiments, and collaborated in the preparation of the manuscript.
F.I. collaborated in the definition and supervision of research tasks, and wrote a substantial part of the manuscript.
K.O. contributed in the critical analysis of the experimental results, and collaborated in the review and writing of the manuscript. 

\subsection*{Funding}

This work has been partially supported by %
Grants PID2021-126576NB-I00 and TED2021-130123B-I00 
funded by MCIN/AEI/10.13039/501100011033 and by {\em “ERDF A way of making Europe”} and
{\em NextGenerationEU/PRT}, and the CM under Grant S2018/TCS-4423.

\subsection*{Availability of data and materials}

Data sharing not applicable to this article as no datasets were generated or analysed during the current study.

\bibliographystyle{sn-basic}
\bibliography{biblio}

\end{document}

%% file: body.tex
\input s1-intro
\input s2-background
\input s3-model_description

\input s4-single_TPU_performance
\input s5-model_segmentation
\input s6-balanced_segmentation

\input s7-conclusions

%% file: s1-intro.tex
\section{Introduction}\label{sec:intro}

Edge Computing aims at bringing computations near to the sensors in Internet of Things (IoT) deployments, in order
to improve latencies, increase security and reduce access cost to datacenters. The convergence
of Edge Computing and Artificial Intelligence (AI) tasks in the so-called Edge-AI paradigm~\cite{Li2019,Ren2022} pursues 
bringing intelligence to edge devices in order to cover a number of applications necessary in many 
IoT scenarios (object detection for smart cameras~\cite{James2019}, smart city applications~\cite{Thalluri2021}, healthcare~\cite{Alshehri2021ACS}
or autonomous driving~\cite{McEnroe2022}), among others), bringing all the benefits of Edge Computing to the AI arena.

The ever increasing necessities of performance for Edge-AI has entailed the emergence of a plethora of 
domain-specific architectures (DSAs) in the form of ASICs (Application-Specific Integrated Circuits) that address
the efficiency problem created by the huge power requirements of general-purpose architectures (e.g. multi-core CPUs or GPUs), in many cases unfeasible
in such scenarios, while still meeting the performance requirements. 
Among them, modern DSAs designed for Machine Learning (ML) primitives such as the Intel NCS~\cite{movidius_tech}, or the Google Edge TPU~\cite{coral_tech} have been recently introduced as an appealing trade-off between performance energy 
efficiency and flexibility for Edge-AI tasks. 

However, DSAs devoted to Edge-AI still suffer from limited performance compared with their general-purpose
counterparts, that ultimately limit their exploitation for compute-intensive tasks. In this work, we investigate
on the implications of using multiple DSAs in a collaborative fashion, taking the Edge TPU as the target architecture.
Employing multiple Edge TPUs to solve a common problem solves the performance problem, while increasing 
energy efficiency, also solving one of the main problems of such architectures: the scarce amount of on-chip
memory per accelerator, that ultimately limits the performance for inference on neural networks with a large memory footprint. 

Specifically, we focus on a setup composed by a PCIe card equipped with eight Edge TPUs that, by segmenting and
pipelining the execution of models, can collaboratively work to solve the problem. The use of multiple TPUs, beyond
the obvious potential performance improvement, alleviates the lack of on-chip memory by spreading the weights of
models across devices, reducing unnecessary host-to-device data transfers and hence further improving performance. 

Model segmentation across devices, however, poses a number of challenges in order to balance the workload in a 
proper way, optimizing resource usage and improving performance. We provide new mechanisms for model segmentation
that improve the ones available in the compiler offered by the vendor, balancing the workload across devices and 
improving the attained performance for a number of representative, real-world models. 

The contributions of the paper can be summarized as:

\begin{itemize}
\item We perform a detailed assessment of the performance obtained for inference processes for synthetic convolutional neural networks (CNNs), observing the correspondence between model size and time-to-solution on a single TPU. 
\item We identify the on-chip memory management and workload balancing problems exhibited by the model segmentation capabilities of the vendor's compiler when targeting multi-TPU processing of CNNs.
\item We propose a profile-based segmentation scheme that extends and improves the segmentation strategy performed by the Edge TPU compiler, and evaluate its benefits on a number of real-world CNNs.
\item We further refine the basic segmentation strategy to balance the workload exposed to each Edge TPU, improving the performance obtained for the compiler's segmentation and also our basic segmentation approach.
\end{itemize}

The observed results reveal high performance gains, doubling the performance of that observed for the compiler's segmentation on many models, and even attaining super-linear speedup when spreading across multiple Edge TPUs.
All performance results for our balanced segmentation have been evaluated for a set of real-world, widely used CNN models. We believe, however, that the strategy is general enough and can be extended to other models and application scenarios.

The rest of the paper is structured as follows.
Section~\ref{sec:background} provides a description of the Edge TPU architecture, and an overview of the state of the art in performance evaluation of TPUs.
Section~\ref{sec:models} describes in detail the CNN models (both synthetic and real-world) used throughout the
paper, and characterizes them in terms of size and number of multiply-accumulate (MAC) operations.
Section~\ref{sec:single_tpu_analysis} provides a preliminary performance evaluation on a single TPU in terms of performance and memory usage for both synthetic and real-world models, identifying the main bottlenecks of the architecture that will be addressed by our segmentation strategies.
Section~\ref{sec:segmentation_multitpu} evaluates the segmentation capabilities of the Edge TPU compiler (\SEGMCOMP) and delves into the necessary mechanisms to provide an optimized profile-based segmentation (\SEGMPROF).
Section~\ref{sec:balanced_segm} provides a refined strategy (\SEGMBALANCED) to alleviate the work unbalance observed in the previous proposals, and assesses the performance obtained compared with previous model segmentation approaches.
Section~\ref{sec:conclusions} closes the paper with some final remarks.

%% file: s2-background.tex
\section{Background}\label{sec:background}

\subsection{The Edge TPU architecture}\label{sec:background:architecture}

\begin{figure}
    \centering
    \centering
    \includegraphics[width=0.7\textwidth]{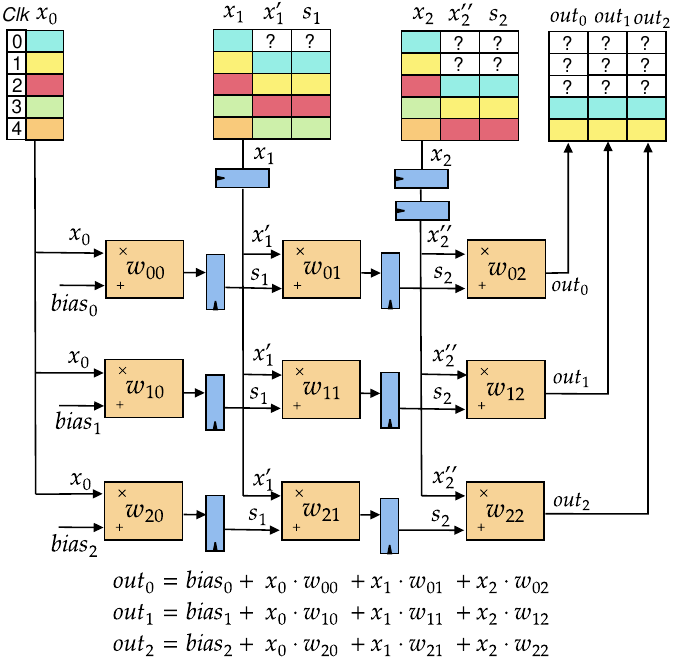}
    \caption{Example of a 3x3 systolic array and the cycle-by-cycle data flow through the chains.}
    \label{fig:systolic_matrix_example}
    \vspace{-15pt}
\end{figure}

The main inference operation in a neural network is the scalar product between input vectors and weight vectors (other operations such as evaluating on the activation function are much less expensive). To speed them up, TPUs include chains of multiply-sum cells~\cite{coral_tech}. In each cell, the product of a weight is calculated by its corresponding component of an input vector, the result is added to the cumulative product of the previous components (received from the previous cell) and propagated forward. These chains are segmented by registers so that the products of different input vectors can be run in parallel (several inferences can be made with the same neural network at the same time). When the size of the vectors exceeds the size of the chain, they are divided into fragments whose scalar products can also be calculated in parallel within a chain (in this case, the partial accumulations of each fragment are reduced to a single scalar at the output of the chain). In addition, the same inputs can be multiplied simultaneously by other weight vectors in other chains. For example, each chain can perform the product of the inputs of a layer by the weights of a neuron and several chains can simultaneously compute several neurons. This multi-chain structure constitutes a matrix of cells known as ``systolic array'' due to the thrust of the data with the clock pulse (similar to the systolic thrust of the blood in the heart). Figure~\ref{fig:systolic_matrix_example} shows an example of a $3\times3$ systolic matrix that calculates the scalar products of several inputs $(x_0, x_1, x_2)$ (each identified with a colour) by the weights $(w_{i0}, w_{i1}, w_{i2})$ of 3 neurons $n_i$ ($i \in \{0,1,2\}$). The propagation of each input cycle by cycle through the chains is indicated by the colours.

Although Google has not disclosed some of the Edge TPU's specifications, there are well-founded estimates that it incorporates a $64\times64$ systolic matrix running at a maximum frequency of 480 MHz. This is consistent with the 4 TOPS peak performance specified in its datasheet
($64 \cdot 64 \text{ cells/cycle } \cdot 2 \text{ ops/cell } \cdot 480 \cdot 10^6 \text{ cycles/s } \simeq 4 \cdot 10^{12} \text{ ops/s}$). In addition, as it is aimed at low power consumption environments, the device uses only 2W of power at maximum performance, offering an energy efficiency of up to 2 TOPS/W. This efficiency is achieved because all calculations in the systolic array are performed with integer arithmetic and multiplications are done with reduced 8-bit precision. This has many benefits in terms of performance, power consumption and hardware cost, but means that the chip can only be used for inference (higher precision is often required for training). In addition, a quantization process is needed before inference to transform the weights of the model trained with \emph{float32} to \emph{int8}.

On the other hand, the Edge TPU also includes an internal memory of 8 MiB where it stores the instructions (CISC repertoire of very high abstraction level), the inputs/activations and the weights of the model. In our setup, the chip is embedded in an M.2 module that connects via PCIe to a host system that invokes it for inference by providing the instructions and data. The host uses the \emph{edgetpu compiler} to generate the code and adapt the model operators to those implemented in the TPU; the models must depart from a \textit{TFLite} model, and be previously quantized to 8-bit integers.

The use of multiple Edge TPUs for collaborative training can be attained by simply attaching more devices to the
PCIe bus. In our case, the experimental setup is based on a PCIe 3.0
card (ASUS CRL-G18U-P3DF~\cite{asus_datasheet}), equipped with
eight M.2 Edge TPU devices; according to its datasheet, the estimated power consumption of the board is 36 W.


%
%

\subsection{Related work}\label{sec:background:relatedwork}

Since its introduction in 2019, both the Cloud and Edge versions of the
TPU have received a significant attention in the literature, mainly assessing their performance and energy
efficiency for model training and inference, respectively.

Several papers have analyzed the cloud version of the TPU.
Specifically, \cite{Nikolic2022, Raj2020} provide an in-depth survey of the differences and particularities
of CPUs, GPUs and (Cloud) TPUs for different tasks related with deep learning.
The use of multiple cloud TPUs to scale training tasks has also been deeply 
studied~\cite{kumar2021exploring, DSA_supercomp}.
The designers of the ASIC presented in \cite{Jouppi2017} a detailed description of the
architecture, evaluating its performance and energy efficiency in comparison with other
state-of-the-art architectures.

The evaluation of multi-device inference engines has also received previous attention
in the literature, but mainly in terms of setups featuring multiple GPUs. 
Specifically, multi-GPU setups have been systematically evaluated in terms of 
energy efficiency~\cite{multigpu_energy} an also performance~\cite{multigpu_accel_spark}.
In the context of GPUs, the problem of host-to-device communications due to insufficient storage space for the model is less critical as the size disparity is not as pronounced. The memory of discrete GPUs, typically used for training in data centers, is in the tens of GiB, and it is suitable for most models such as the CNNs we use. Although GPUs in SoCs, more suitable for inference, have smaller memory, it is still far from the constraints of devices like the Edge TPU, and the problems usually lie in the size of the input batches rather than the model~\cite{Guo2019}. Moreover, in this case the GPU memory is usually shared with the CPU and these input-output problems are reduced or disappear. However, this problem becomes significant with recent Large Language Models~(LLMs), whose enormous size necessitates the use of pipeline partitioning with approaches such as AlpaServe~\cite{Zhuohan2023}, PipeDream~\cite{Narayanan2019} or Gpipe~\cite{Huang2019}. These techniques are often employed alongside aggressive model compression methods, such as quantization~\cite{Zhou2018} or pruning~\cite{Ma2013}, that have received more attention in the literature due to the challenge of applying these techniques without losing much accuracy.


In addition, current pipeline partitioning could be quite expensive as they try to approximate the optimal solution very well. For instance, AlpaServe~\cite{Zhuohan2023} profiles the inference time of each possible pair of levels in the DAG associated with the model, resulting in a quadratic number of profiles relative to the model's depth. This means tens of thousands of profiles, which can only be obtained by running inferences on devices such as the Edge TPU. Similarly, PipeDream~\cite{Narayanan2019} has worse than cubic complexity in terms of model depth and reports partitioning times of around ten seconds when run on powerful machines with not very large models. This complexity is manageable in the context of specific DNNs trained over a long period for deployment on the GPU, as the cost is more than amortized with a significant increase in training speed and prolonged use of the model for inference. In fact, costly retraining is also assumed for fine-tuning the model during pruning or quantization~\cite{Kim2023,Renda2020}. In contrast, this is not assumable for deploying only inference at the edge, which usually require more dynamism and present more variety; at the edge it is reasonable to receive many different CNNs for specific tasks (object detection, segmentation, face recognition, etc.), which come from different users and need a fast response~\cite{He2020,Hu2019,Zeng2021}. For us, fast partitioning is crucial to facilitate an efficient, although suboptimal, model inference.

Many other works are oriented to edge-cloud scenarios, where traditionally DNNs are split in two parts between the edge and the cloud with the objective of minimizing the total communication and computation time while guaranteeing data privacy \cite{He2020,Hu2019,Mohammed2020,Li2021,Wu2023}. However, we study the case where there are multiple TPUs within an edge device, so our partitioning model is not limited to two fragments, but can span multiple, and communication delay is much less important. This invalidates the application of algorithms for bipartite partitioning used in previous proposals such as min-cut \cite{Hu2019}.

Few papers address multi-segment partitioning in scenarios similar to ours. Some employ {\em model parallelism}~\cite{Du2021,Mohammed2020,Zhou2023}, which splits different paths in the graph associated with the model by fragmenting the layer tensors. Others use {\em pipeline parallelism}~\cite{Parashar2020,Xiang2019,Zhou2019}, which entails splitting the model at certain depth levels so that entire layers remain within each segment. For devices like the Edge TPU, model partitioning is not feasible due to the lack of support for executing fragmented tensors, as we will see in Section~\ref{sec:memory_usage}. Pipeline partitioning approaches typically aim to enhance performance when there are more devices than models, without accounting for the memory constraints of fitting an entire model into memory. These approaches assume that the execution times of each layer, considered for optimizing partitioning, do not vary with the segment size, which is not pertinent for memory-constrained devices like the Edge TPU. Considering variable execution times as a function of segment size would require profiling all possible partition points, as in~\cite{Zhuohan2023}, which would be prohibitively costly. Therefore, instead of profiling execution times, we will use an intrinsic model parameter (the number of weights by level), which is deduced from our performance study as a good indicator to mitigate the bottleneck of host-device communications when the fragments do not fit in the devices' memory.

The Edge version of the TPU has received less attention in the literature~\cite{Cass2019}.
\cite{Murshed2022} considers the Edge TPU as one of the testsbeds to evaluate ASICs
for image processing tasks.
\cite{Kang2022} evaluates the Edge TPU for object detection activities at the edge.
\cite{Antonini2019} provides a comparative study of different edge accelerators, including
the Edge TPU, for personal sensing applications.
\cite{Seshadri2021,Sun2021} generalize the study by covering a number of DL 
models for the performance evaluation of the Edge TPU. The previous works, however, provide
insights for specific (not synthetic) deep neural networks and/or applications, without
further insights for generalizing the conclusions to other (existing or non-existing) models; 
none of them covers the use of multiple TPUs in the evaluation.

The Edge TPU has also been included as a member of the family of devices
capable to accelerate Edge-AI tasks by means of benchmarks. \cite{Varghese2021}
includes the device as one of the target architectures appealing for state-of-the-art
on edge performance benchmarks.
MLPerf
is an initiative to design portable benchmarks and
to evaluate them on different architectures. MLPerf defines a
set of models that are evaluated for training (mainly using high performance
architectures, e.g. GPUs) and also for inferencing. In the latter case,
{\em MLPerf Inference} includes different variants for datacenter, mobile,
tiny and edge devices. Although the Edge TPU has not been
included in any results for the edge MLPerf inference benchmarks,
\cite{Libutti2020} performed a detailed study of its performance
for a subset of the models used in the benchmark for the USB version
of the device. 

The use of multiple Edge TPUs combined with profiled model segmentation was previously
addressed in~\cite{Villarrubia2023}. This paper extends the previous study by providing
a more refined balanced segmentation approach, and by evaluating real-world models, beyond
the synthetic evaluation performed in the original work. The obtained results support and extend, for
widely-used models the efficiency of multi-TPU setups for inference in large CNNs.

%% file: s3-model_description.tex
\section{Selected CNN models\label{sec:models}}


When mapped to a systolic array with limited on-chip memory such as the Edge TPU, we can identify three general features with
impact in performance in CNNs, namely:

\begin{itemize}
 \item {\em Workload}: number of multiply-accumulate operations (\textit{MACs}) per forward pass for inference, that will ultimately impact the arithmetic intensity of the procedure.
 \item {\em Size}: number of weights and hence amount of memory occupied by the model in on-chip memory.
 \item {\em Architecture}: number, characteristics, type and combination of layers in the model.
\end{itemize}

In CNNs, the workload (number of MACs) and size (number of weights) are intimately related dimensions, as is their
impact in ALU utilization and memory footprint, respectively. It is hence natural that a systematic assessment
of performance and memory utilization for CNN inference requires evaluating models featuring a gradual increase in both dimensions.
For that end, we will commence our evaluation using {\em synthetic} models with similar architecture, generated in a parametric
fashion in order to gradually increase both dimensions simultaneously, as depicted in Section~\ref{sec:synthetic_models}.

The use of synthetic models allows a parametric evaluation and eases the extraction of remarkable insights; however, it
does not evaluate the impact of the network architecture in performance when mapped to the Edge TPU, and it is not easily
extended to real-world scenarios. To address these issues,
we will also consider real-world CNN models as described in Section~\ref{sec:real_world_models} to extend and consolidate the conclusions of the
parametric study performed with synthetic models. Actually, the impact in performance of our segmentation approach applies to
a greater extent for real models.

\subsection{Synthetic CNNs}\label{sec:synthetic_models}

To obtain a gradual growth in the size of our synthetic models we have used a family with a progressive increase in the number of trainable parameters (weights). Our synthetic models are CNNs composed by $L$ convolution layers with $f$ filters of size $F_w \times F_h$ per layer, applied with stride $1$ and padded with zeros. In these models, the input layer consists of $C \cdot f \cdot F_w \cdot F_h$ parameters, where $C$ is the number of input channels. For the remaining layers, the number of input channels matches the number of filters $f$ of the previous layer, yielding $f \cdot f \cdot F_w \cdot F_h$ parameters in each one. Summarizing, the number of parameters as a function of the number of filters per layer is $\#\text{params}\,(f) =  F_w \cdot F_h \cdot f \left(C + f \cdot (L-1)\right)$, that grows linearly with $f$ if $L= 1$ or quadratically if $L > 1$. Thus, we can gradually increase the amount of parameters by increasing the number of filters $f$ with a fixed value for the rest of the parameters.

In fact, this way of generating the models also produces a progressive increase in the number of MACs, i.e. in the model workload (the number of MACs is the number of parameters multiplied by the input dimensions $W \times H$, which are constant and equal for all layers due to padding with zeros).



The family of synthetic models used throughout the paper is obtained by taking $L = 5$, $C = 3$, $W \times H = 64 \times 64$, $F_h\times F_w = 3\times 3$, and varying $f$ between $32$ and $1152$ with step $10$. 
This set of models is useful for illustrative purposes in terms of performance penalties that will be discussed 
in Section~\ref{sec:single_tpu_analysis}
and at the same time makes it possible to apply a naive segmentation strategy described in 
Section~\ref{sec:segmentation_multitpu}, due to the reduced number of layers. The purpose of this selection is to facilitate a parametric study in which relevant performance aspects are clearly more visible, that can afterwards be also verified for real-world CNN models.

\subsection{Real-world CNNs\label{sec:real_world_models}}

To develop our experiments and implementations we have used TensorFlow Keras, so it was especially convenient for us to use the models provided by this API\footnote{\url{https://keras.io/api/applications/}}. These models are the main CNNs used for image classification and form the basis of popular networks for other image problems such as object detection or pose estimation.

\input{real_CNNs_description}

We decided to discard the Keras models that are too large and would never be used in lite devices such as the Edge TPU. As our study evaluates the segmentation of models among several TPUs in such a way that the models can be stored entirely in their memories, we discarded those that would require more than 8 Edge TPUs, which is the number used to evaluate our proposal. For example, \textsc{NASNetLarge} was not used as it occupies $\sim \! 88.9$ MiB in quantized TFLite format and would require at least $12$ Edge TPUs to avoid host memory usage ($\lceil88.9/8\rceil = 12$, because each Edge TPU has a memory of $8$ MiB). In addition \textsc{EfficientNet} models we discarded as they incorporate dynamic tensors that are not supported by the TFLite interpreter. Instead, we used the corresponding lite versions developed by TensorFlow\footnote{\url{https://github.com/tensorflow/tpu/tree/master/models/official/efficientnet/lite}}. Table~\ref{tab:real_CNNs_description} reports some characteristics of these real CNNs that may be relevant for this study.

%% file: real_CNNs_description.tex
\begin{table}[t]
\centering
\caption{Real-world CNNs with their number of parameters, number of MACs, depth and size using 8-bit integer quantization and TFLite format.}
\label{tab:real_CNNs_description}
\begin{tabular}{|l|c|c|c|c|}
\hline
\rowcolor[HTML]{EFEFEF} 
\multicolumn{1}{|c|}{\cellcolor[HTML]{EFEFEF}Model name} &
  \begin{tabular}[c]{@{}c@{}}Params\\ (millions)\end{tabular} &
  \begin{tabular}[c]{@{}c@{}}MACs\\ (millions)\end{tabular} &
  Depth\footnotemark[1] &
  \begin{tabular}[c]{@{}c@{}}Quantized size\\ (MiB)\end{tabular} \\ \hline
Xception           & 22.9 & 8363  & 81  & 23.07 \\
ResNet50           & 25.6 & 3864  & 107 & 25.07 \\
ResNet50V2         & 25.6 & 3486  & 103 &  25.12 \\
ResNet101          & 44.7 & 7579  & 209 & 42.88 \\
ResNet101V2        & 44.7 & 7200  & 205 & 43.96 \\
ResNet152          & 60.4 & 11294 & 311 & 59.41 \\
ResNet152V2        & 60.4 & 10915 & 307 & 59.53 \\
InceptionV3        & 23.9 & 5725  & 189 & 23.22 \\
InceptionV4        & 43.0 & 12276 & 252 & 40.93 \\
MobileNet          & 4.3  & 568   & 55  &  4.35 \\
MobileNetV2        & 3.5  & 300   & 105 & 3.81 \\
InceptionResNetV2  & 55.9 & 13171 & 449 & 55.36 \\
DenseNet121        & 8.1  & 2835  & 242 & 8.27 \\
DenseNet169        & 14.3 & 3361  & 338 & 14.02 \\
DenseNet201        & 20.2 & 4292  & 402 & 19.71 \\
NASNetMobile       & 5.3  & 568   & 389 & 6.11 \\
EfficientNetLiteB0 & 4.7  & 385   & 208 & 5.00 \\
EfficientNetLiteB1 & 5.4  & 600   & 208 & 5.88 \\
EfficientNetLiteB2 & 6.1  & 859   & 208 & 6.58 \\
EfficientNetLiteB3 & 8.2  & 1383  & 238 & 8.83 \\
EfficientNetLiteB4 & 13.0 & 2553  & 298 & 13.87 \\ \hline
\end{tabular}

\footnotetext[1]{The depth is the maximum number of layers between any input and any output.}

\end{table}

%% file: s4-single_TPU_performance.tex
\section{Analysis of inference performance and memory usage on a single TPU\label{sec:single_tpu_analysis}}

In this section, we evaluate the performance and memory usage of both synthetic models and real-world models as 
described in the previous sections. In all cases, {\tt int8} quantized models have been deployed, and inference
has been carried out via the Python API of TFLITE on CPU, and the Edge TPU delegate in the case of the Edge TPU.
The standard operating frequency of the Edge TPU was selected in order to reduce undesired thermal-aware frequency-throttling
effects.
50 repetitions of each inference were averaged to report performance results, and the complete execution time (including
data transfers) is reported in all cases\footnote{The described experimental conditions are also used throughout the rest of the paper.}.

\subsection{Performance evaluation}

\begin{figure}[ht]
\centering
    \includegraphics[width=0.8\textwidth]{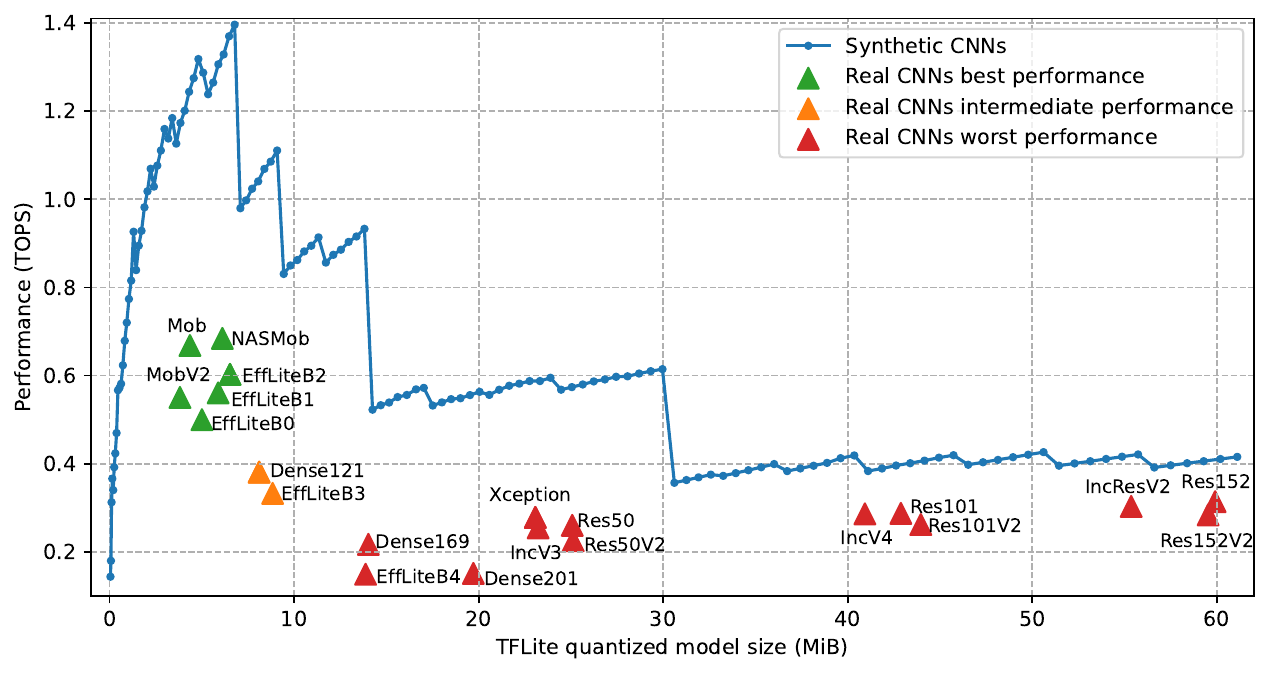}
\caption{Average performance of inference for synthetic and real models (in TOPS) after 50 repetitions using batch size 1, as a function of the model size.}
\label{fig:performance}

\includegraphics[width=0.8\textwidth]{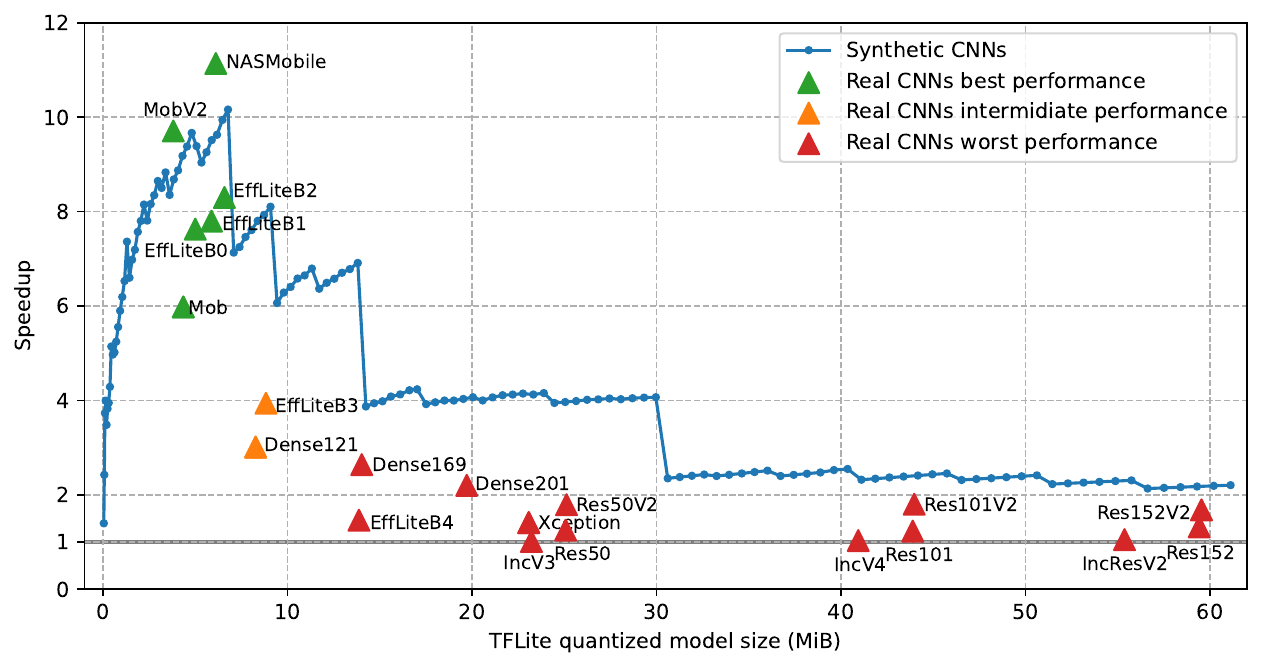}
\caption{Speedups of synthetic and real model inference on Edge TPU vs. Intel \mbox{i9-9900K} with $8$ threads using batch size 1.}
\label{fig:speedup_TPU_vs_CPU}
\end{figure}

Let us start by assessing the inference performance of the selected CNN models on a single
Edge TPU. 
Figure~\ref{fig:performance} reports the performance of each model as a function of its size in
terms of TOPS ($10^{12}$ {\tt int8} operations per second). The results of the synthetic models are joined by a blue curve to highlight their trend, and the results of the real models are grouped into 3 clusters by colors based on their performance (see plot legend). 
From these results, a number of specific insights can be extracted:

\begin{itemize}
    \item The performance of synthetic models clearly evolves in a stepped fashion. Each step consists of a gradual increase in performance until a sharp drop occurs. The performance improvement at each step is reasonable as the workload of the models increases and the systolic array pipeline is better amortized (more cycles running with maximum parallelism). On the other hand, the stepped behaviour is caused by longer waits to load data into the systolic array; this observation will be sustained with data in the next section.

    \item Although the real models are more sparse in terms of observed performance, they also seem to follow a similar stepped behavior. Models have been classified into three different groups, colored in green, orange and red, respectively. The models in the green group show the best performance by far and are the smallest in size. The two models colored in orange (slightly larger in size) exhibit worse performance. Finally, there is a large group of models colored in red, with much worse performance than the previous ones and larger sizes. The models colored in red exhibit a decrease in performance with increasing size similar to those observed for the synthetic model steps.

    \item The performance of both model families is far from the theoretical peak ($4$ TOPS). The runs are heavily penalized by performance drops, but even before the drops, the models perform well below the ideal: $1.4$ TOPS in synthetic models and only $0.6$ TOPS in real models (green group). Peak performance would be obtained if there were no fill latencies in the systolic array and, especially, no stalls waiting for data. The observed results suggest that executions are highly memory bound. This also explains the better performance of synthetic models, which only use convolutions, versus real models, that also use layers of lower arithmetic intensity (e.g., fully connected).
\end{itemize}

Putting the performance of the Edge TPU into perspective, and although the device is not used at its full potential, its performance is relatively good in absolute terms. Figure~\ref{fig:speedup_TPU_vs_CPU} shows the inference speedup of our models on the Edge TPU against a general-purpose CPU: an 8-core Intel i9-9900K CPU, using 8 threads and running at a nominal frequency of 3.6 GHz. With the synthetic models, a speedup of $10 \times$ is reached at the end of the first step and, despite successive performance drops, the speedup remains above $2 \times$ for the rest of models. With the real models, the improvement is even larger in the best-performing group (green color) with speedups close to $12\times$ in some cases. The improvement for the rest of the models (orange and red colors) are less dramatic. However, the Edge TPU is never slower than the multi-core CPU. In the following, we focus on understanding and mitigating the performance problems of these models.

\subsection{Memory usage}\label{sec:memory_usage}

Motivated by the previous observations, we proceed by analyzing the executions in terms of memory accesses. As of today, and to the best of our knowledge, there are no profiling tools to obtain this type of metrics. Nevertheless, the model compiler generates a report that includes the amount of host and device memory used by the Edge TPU to store the weights. This information is a good indicator of the cost associated with memory operations, since weight read operations are dominant in the inference process in terms of execution time (the tensors associated with reading inputs or writing outputs are considerably smaller).

\begin{figure}[t]
\centering
\includegraphics[width=0.8\textwidth]{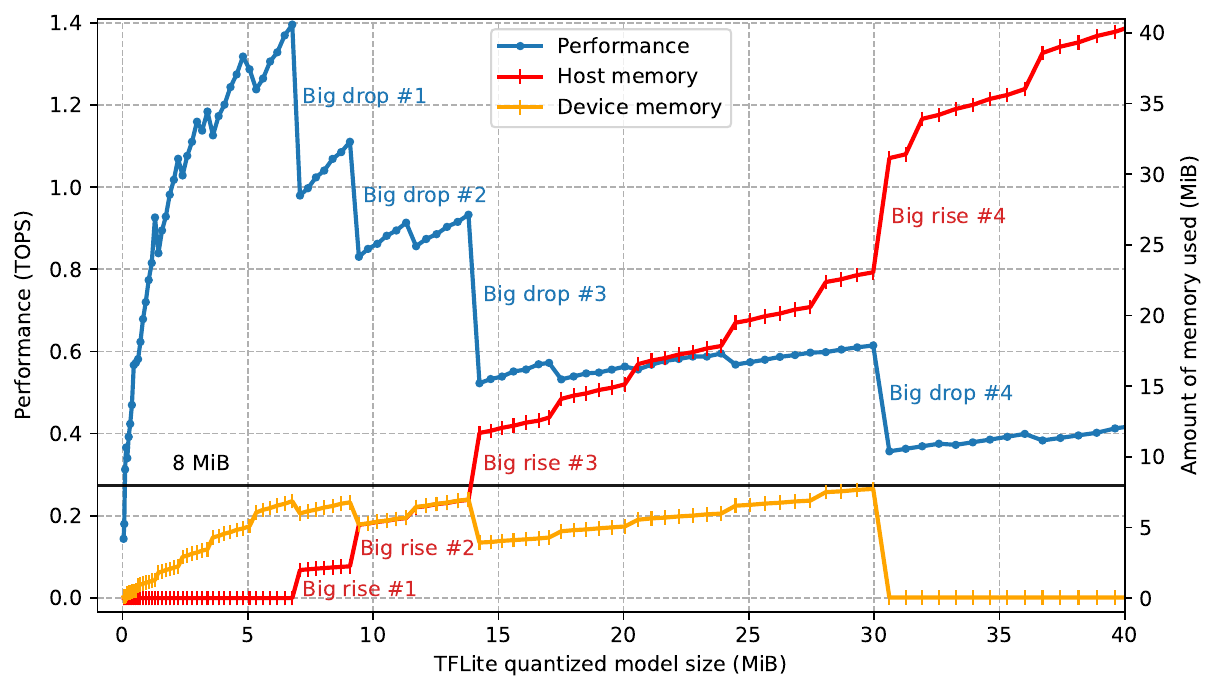}
\caption{Performance of the synthetic models (blue curve associated with the left vertical axis) and their device and host memory usage (yellow and red curves associated with the right vertical axis).}
\label{fig:memory_usage}
\end{figure}

Figure~\ref{fig:memory_usage} shows the performance of the synthetic models along with its device and host memory usage. Memory usage perfectly explains the big performance drops that were previously observed (annotated in blue in the figure). At each step, the device memory usage grows progressively until it almost reaches the available size ($8$ MiB). Then, a big performance drop occurs, matching with a drastic rise in host memory usage (annotated in red in the figure). This happens when part of the model cannot be stored on the device and commences to be stored in the host memory. The overhead of loading these weights is the primary source of the observed performance losses.

Between big performance drops, small performance drops are also observed, coinciding with small sharp increases in host and device memory usages. Even though this information is not disclosed, a reasonable assumption is that this effect is caused by the compiler padding the tensors with zeros to make their sizes multiple of the dimensions of the systolic array. So, performance drops would occur because large tensors of the model just exceed a multiple of the array dimensions and have to be filled with many zeros that consume time in useless operations. Mitigating this issue is not straightforward, as it is directly managed by the compiler, but its impact is minimal compared to the overhead of loading host memory weights.

Host memory usage occurs in abrupt steps because the neural layer is the minimal storage unit: the Edge TPU compiler stores all weights of a layer in the same memory space. To reach this conclusion, we must consider that our synthetic models have $4$ large layers with $f^2$ kernels of the same dimensions ($f$ filters over $f$ input channels), and a much smaller input-layer with just $3f$ kernels ($f$ filters over $3$ input channels). Table~\ref{tab:drops_synthetic_models} shows that, in the first big performance drop, the host memory goes from not being used to storing $25\%$ of weights, since it stores one of the four large layers. With the second drop the host memory starts to save $50\%$ corresponding to two large layers, leaving the other two in the device. Similarly occurs with the third and fourth big drops. Theoretically, the tensors could be split to store only the strictly necessary part of the model in the host, but the compiler proceeds by storing full tensors, presumably for easier weight management. This solution should make the same number of memory copies as a storage scheme with a finer granularity, but with more data in each one. 
\input{drops_synthetic_models}

\input{real_models_memory_usage}

On the other hand, the memory usage of the real models (see Table~\ref{tab:real_models_memory_usage}) sustains that the penalties for using host memory are the cause of the performance losses. We observe that the models of the best-performing group (green color) are the only ones that do not require host memory. The orange models require a small amount of host memory compared to the red ones (around $2$ MiB vs. tens of MiB) and thus their performance is a bit better. Among the red models there are also big differences in the amount of host memory used (from $8.59$ MiB to more than $50$ MiB), but their performances were similar. In fact, we saw a slight performance improvement by increasing their model sizes in Figure~\ref{fig:performance}, and it seems that host communications overhead is saturating for them in some way.

In summary, communications with the host (in our case, through the PCIe bus) are a non-negligible bottleneck that should be reduced or avoided. In the following we propose strategies to reduce them through model segmentation.

%% file: drops_synthetic_models.tex
\renewcommand{\arraystretch}{1.25}
\begin{table*}[ht]
\centering
\caption{Device and host memory usage of synthetic models before and after each performance drop.}
\label{tab:drops_synthetic_models}
\begin{tabular}{|c|c|c|c|}
\hline
\rowcolor[HTML]{EFEFEF} 
Big drop               & 
\begin{tabular}[c]{@{}c@{}}Model size \\ (MiB)\end{tabular}
& \begin{tabular}[c]{@{}c@{}}Device memory \\ (MiB)\end{tabular}
& \begin{tabular}[c]{@{}c@{}}Host memory \\ (MiB)\end{tabular}  \\ \hline
                    & 6.86 & 6.86 ($100\%$) & 0 ($0\%$)\\
\multirow{-2}{*}{\#1} & 7.98 & 5.99 ($\sim 75\%$) & 1.99 ($\sim 25\%$) \\ \hline
                    & 9.03 & 6.78 ($\sim 75\%$) & 2.25  ($\sim 25\%$)\\
\multirow{-2}{*}{\#2} & 10.41 & 5.21 ($\sim 50\%$) & 5.19 ($\sim 50\%$) \\ \hline
                    & 13.94 & 6.98 ($\sim 50\%$) & 6.95  ($\sim 50\%$) \\
\multirow{-2}{*}{\#3} & 15.62 & 3.93 ($\sim 25\%$) & 11.69  ($\sim 75\%$) \\ \hline
                    & 30.79 & 7.73 ($\sim 25\%$) & 23.06  ($\sim 75\%$) \\
\multirow{-2}{*}{\#4} & 31.18 & 0.04 ($\sim 0\%$) & 31.14  ($\sim 100\%$) \\ \hline
\end{tabular}%
\end{table*}
\renewcommand{\arraystretch}{1}

%% file: real_models_memory_usage.tex
\begin{table*}[ht]
    \caption{Amount of memory used by the real models indicating its color group.}\label{tab:real_models_memory_usage}

    \small

    \begin{minipage}[t]{0.49\textwidth}
      \centering
      \begin{tabular}{|
        >{\columncolor[HTML]{FAB5BE}}l |c|c|}
        \hline
        \multicolumn{1}{|c|}{\cellcolor[HTML]{EFEFEF}Model} & \cellcolor[HTML]{EFEFEF}\shortstack{Device\\(MiB)} & \cellcolor[HTML]{EFEFEF}\shortstack{Host\\(MiB)} \\ \hline
        Xception                            &  6.22 &  17.72 \\
        ResNet50                            &  7.14 &  17.54 \\
        ResNet50V2                          &  7.14 &  17.96 \\
        ResNet101                           &  7.23 &  35.90 \\
        ResNet101V2                         &  7.23 &  36.83 \\
        ResNet152                           &  7.31 &  51.04 \\
        ResNet152V2                         &  7.31 &  52.42 \\
        InceptionV3                         &  6.13 &  17.97\\
        InceptionV4                         &  6.13 &  36.30 \\
        \cellcolor[HTML]{D8F8B8}MobileNet   &  4.12 &  0 \\
        \cellcolor[HTML]{D8F8B8}MobileNetV2 &  3.88 &  0 \\ \hline
        \end{tabular}
    \end{minipage} %
    \hfill
    \begin{minipage}[t]{0.49\textwidth}
      \centering
        \begin{tabular}{|l|c|c|}
        \hline
        \rowcolor[HTML]{EFEFEF} 
        \multicolumn{1}{|c|}{Model} & \shortstack{Device\\(MiB)} & \shortstack{Host\\(MiB)} \\ \hline
        \cellcolor[HTML]{FAB5BE}InceptionResNetV2           &      6.52        &      49.61      \\
        \cellcolor[HTML]{FCDFBD}DenseNet121                 &      7.04        &      2.98      \\
        \cellcolor[HTML]{FAB5BE}DenseNet169                 &      7.04       &      8.59      \\
        \cellcolor[HTML]{FAB5BE}DenseNet201                 &      7.04        &     15.17       \\
        \cellcolor[HTML]{D8F8B8}NASNetMobile                &      6.31        &      0      \\
        \cellcolor[HTML]{D8F8B8}EfficientNetLiteB0          &      5.32       &      0      \\
        \cellcolor[HTML]{D8F8B8}EfficientNetLiteB1          &      6.33        &      0      \\
        \cellcolor[HTML]{D8F8B8}EfficientNetLiteB2          &      7.38        &      0      \\
        \cellcolor[HTML]{FCDFBD}EfficientNetLiteB3          &      7.56        &      2.18      \\
        \cellcolor[HTML]{FAB5BE}EfficientNetLiteB4          &      7.56        &      7.93    \\ \hline
        \end{tabular}
    \end{minipage}
\end{table*}

\setlength{\tabcolsep}{6pt}

%% file: s5-model_segmentation.tex
\section{Model segmentation for multi-TPU inference\label{sec:segmentation_multitpu}}

\subsection{Overall goal and pipeline implementation}

\begin{figure}[h]
    \centering
    \includegraphics[width=0.9\textwidth]{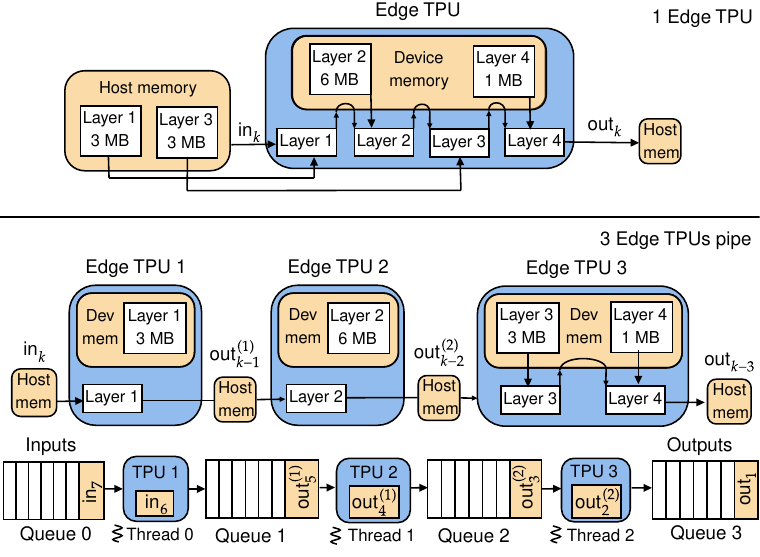}
    \caption{Top: Single TPU execution of a model with layers stored on the host. Bottom: Pipelined execution of the model, segmented into 3 TPUs, without layers stored in host memory.}
    \label{fig:segmentation_TPUs_example}
\end{figure}

Our proposal to reduce host memory usage (and hence the associated penalty in performance), 
is to segment the models and distribute the fragments across several Edge TPUs. 
The proposed strategy is to expand the effective device memory space by aggregating 
multiple TPUs so that less host memory is needed and host-to-device communications are reduced. 
In order to run inferences, the outputs of each segment are used as inputs to the TPU 
that contains the next one (see Figure~\ref{fig:segmentation_TPUs_example}). 
Although these communications are also carried out via host memory, they can be cost-effective 
because the number of intermediate outputs that are transmitted among TPUs is much smaller than 
the amount of the weights we avoid sending (almost $8$ MiB for each additional TPU).

Proceeding in this manner, we build a pipeline of devices that allows parallel execution 
of several inputs at different stages. Thus, we can better amortize the executions of a 
multi-input batch instead of individual inputs. 
The latency constraints of edge processing do not allow waiting many data read periods to 
accumulate a batch, but it is common to have several data sources gathering data at once that 
allow forming a small batch for each read period (e.g., many cameras for object detection or
many source of telemetry data). 
To implement the pipelined execution, we deploy a host thread per Edge TPU that is in charge of
handling it, and a queue (implementing thread-safe mechanisms) on the host to communicate 
intermediate results among devices (see Figure~\ref{fig:segmentation_TPUs_example}).

\subsection{Compiler-based segmentation ({\sc Segm\_Comp})}\label{sec:compiler_segmentation}

The Edge TPU compiler includes a tool for model segmentation\footnote{\url{https://coral.ai/docs/edgetpu/compiler/\#model-segmentation}}. By leveraging a specific compilation option, the user can indicate the number of segments to form, and the compiler is in charge of
producing an executable file for each one. Then, the segmented model can be run in multiple TPUs as segments according to the pipeline implementation explained above. 
In the following, we refer to this compiler-based segmentation strategy as \SEGMCOMP.
Next, we evaluate this segmentation approach on a $15$-input batch, so that the execution takes advantage of the parallelism of the pipeline. An analysis of single-input executions can be found in our previous paper~\cite{Villarrubia2023}.

\subsubsection{Analysis of {\sc Segm\_Comp} for synthetic models}

\begin{figure}[h]
    \centering
    \includegraphics[width=0.8\textwidth]{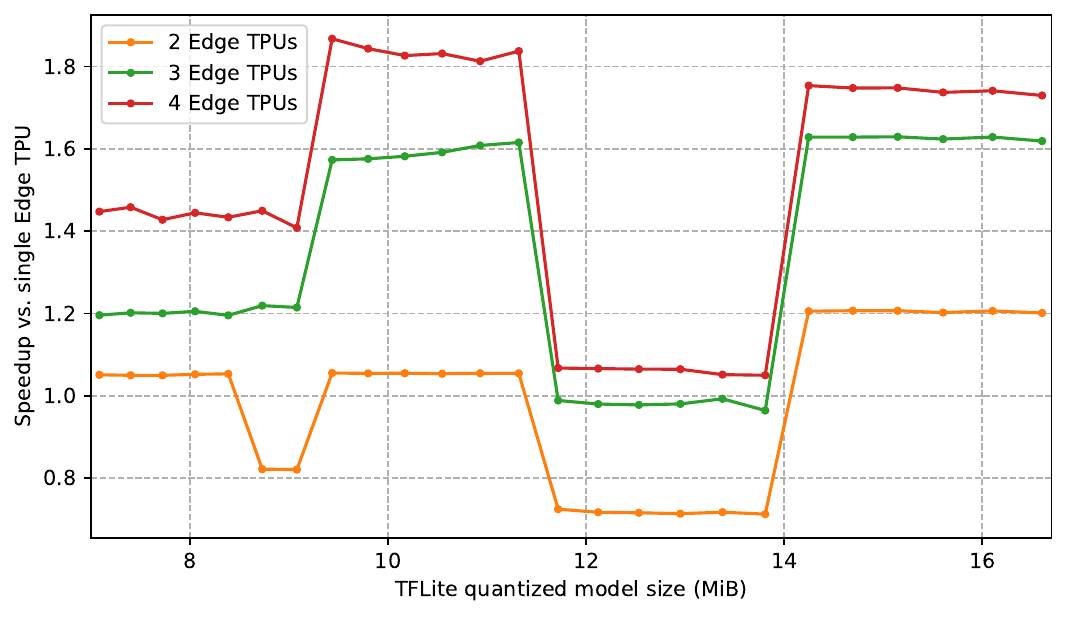}
    \caption{Speedup of synthetic models using \SEGMCOMP, segmented into $2$, $3$ and $4$ TPUs run on a $15$-input batch, versus execution on a single TPU.}
    \label{fig:speedup_comp_seg_vs_1_TPU}
   \vspace{-15pt}
\end{figure}

Figure~\ref{fig:speedup_comp_seg_vs_1_TPU} shows the inference speedup of our synthetic models\footnote{The synthetic models used are those that require host memory (after the first performance drop), and can leverage the extra memories because their layers occupy less than $8$ MiB (before the fourth performance drop). A situation with layers occupying more than $8$ MiB was illustrative above, but it is purely synthetic and unimportant because it does not occur in real models. when using the \SEGMCOMP strategy to segment into two, three and four fragments, versus running the entire model on a single TPU. We note that speedups are well below the number of TPUs used, with a maximum of $1.8 \times$ using $4$ TPUs. This is actually a disappointing result, since by simply replicating the model on the TPUs (i.e. exploiting {\em model parallelism}) and partitioning the input batch (i.e. exploiting {\em data parallelism}), we would potentially obtain a more efficient execution (yielding speedups closer to the number of TPUs employed). In fact, for models between 12 and 14 MiB we see that \SEGMCOMP is even slower or almost the same as with a single TPU. This is because it makes an inefficient partition that underutilises the memory of one of the TPUs and needs to leave a layer on the host, just like when using a single TPU. Thus, due to the huge host-to-device communication cost relative to the inference time, the segmented execution takes about the same but with the additional cost of pipeline communications. This will become clearer below through the results of Table~\ref{tab:mem_usage_synthetic_4_seg}.}
We also appreciate sudden speedup drops even for $4$ TPUs, which should be enough to store the $4$ large layers of these models (one layer per TPU). Therefore, it is clear that the segmentation carried out by the compiler is not ideal and the host memory is still in use.

Table~\ref{tab:mem_usage_synthetic_4_seg} provides a detailed analysis of the reports on device and host memory usage
returned by the compiler for a set of synthetic models segmented in $4$ parts. The analysis of this report clarifies 
the main drawback of \SEGMCOMP, as the compiler produces dramatically unbalanced segments in terms of size.
%
%
In this example, observe how the memory of the first TPU is under-utilized with a segment that occupies very little size (less than $0.1$ MiB), but the memory of the fourth TPU is over-utilized with a huge segment. This situation causes the fourth TPU to eventually have to use host memory. In the larger models, performance is not as expected due to host memory usage that could be avoided with a more balanced segmentation. However, the other models also do not perform as expected because an unbalanced partitioning of the size is obviously linked to an unbalanced workload distribution, which negatively affects the performance of the execution pipeline.

Our experience is that the compiler balances the number of layers in the segments, but not the number of model parameters in each as the documentation indicates. In the example shown in Table~\ref{tab:mem_usage_synthetic_4_seg} we can see that the 5 layers of the model are distributed as 1-1-1-2, although the first layer is very small and the last two quite large, even when necessary. A 2-1-1-1 distribution would have followed better in this case. This has also been observed with real models, although it is more difficult to illustrate so clearly due to their dimensions.

\input{mem_usage_synthetic_4_seg}

We acknowledge that our synthetic models feature four large layers and one small layer, and each layer is stored completely in one of the memory spaces. Thus, based on the data in Table~\ref{tab:mem_usage_synthetic_4_seg}, it seems clear that the first fragment contains only the small layer, the second and third fragments contain one large layer each (same large usage amounts), and the fourth fragment contains the remaining two large layers (double the size of the second and third). When the fourth TPU uses host memory, we clearly see that it keeps one of the two large layers of its segment, keeping the other on the device (same host and device memory uses since then). This strategy is obviously improvable by simply making the first fragment contain a large layer next to the small one, which would free the last segment from the layer that forces the use of the host.

\subsubsection{Analysis of {\sc Segm\_Comp} for real models}

Although we cannot analyze the real models to the point of understanding what happens with each layer as
we did in the previous section with synthetic models, we can actually check if the compiler also yields an 
unbalanced segmentation on them. 
For that, we have fragmented each model into enough TPUs to avoid host memory usage provided
\SEGMCOMP could yield an ideally balanced partition: a model occupying $S$ MiB has been fragmented into 
$\lceil S/8 \rceil$ TPUs (as each TPU can store up to $8$ MiB). 
Table~\ref{tab:real_models_comp_segm} shows the host memory usage of the segmented models, 
along with the size difference between the largest and smallest segment ($\Delta_s$) as a metric for segmentation imbalance, 
and the inference speedup against single-TPU execution; the speedup results are given in absolute terms and also normalized
to the number of TPUs (in parenthesis). Regarding these results, a number of insights can be extracted:

\input{real_models_compiler_segmentation}

\begin{itemize}
    \item Although the number of TPUs is sufficient to completely avoid host memory usage, the compiler segmentation fails to do so for seven of the fifteen tested models. Host memory usage is small in two of them (marked in light red), but is quite large 
    in the other five (marked in darker red).

    \item $\Delta_s$ is quite significant in almost all cases (in the order of several MiB). This indicates an unbalanced segmentation that seems slightly larger in the models that still require host memory usage. The two {\sc EfficientNetLite} models are an exception in that the compiler segmentation is sufficiently balanced (marked in green).

    \item Models that still require host memory show a much worse speedup than the number of TPUs. In contrast, models that avoid host memory show a speedup close to the number of TPUs. This result is better than that observed for the synthetic models, but it is also poor: they are justified only by an efficient use of the pipeline, but we are also avoiding costly loads from host memory. It seems that the execution in the pipeline is not efficient due to the detected imbalance. On the {\sc EfficientNetLite} models, which exhibit a more balanced partitioning (lower $\Delta_s$), the normalized speedups are above $1 \times$.
\end{itemize}

\subsection{Improving model segmentation with profiling ({\sc Segm\_Prof})}\label{sec:profiling}

As seen, there is still room for improvement in \SEGMCOMP through a more balanced partitioning of the model size. 
In this sense, we will present our own segmentation scheme in Section~\ref{sec:balanced_segm},
specifically focusing on real models.
Let us first introduce a simpler optimization technique in detail:
run and profile each possible partition in the pipeline of TPUs to profile its 
performance and choose the best one. In the following, we refer to this
profiled-based segmentation as \SEGMPROF.

\SEGMPROF is mainly affordable for shallow models in terms of number of layers. 
Provided we segment the models by separating the layers at a certain depth 
(we will see next that real models present even more options), there are  $\binom{d-1}{s-1}$ possible 
partitions\footnote{It is about splitting $d$ depth levels into $s$ segments. This is equivalent to choosing $s-1$ separators (to form the $s$ segments) among the $d-1$ positions between depth levels. That is, there are $\binom{d-1}{s-1}$ options.}, 
where $d$ is the depth of the model and $s$ the number of segments to be formed. Exploring all of them is possible for shallow networks like our synthetic models, where there are just tens of possibilities ($d=5$ since there are $5$ layers one after the other). However, it has an excessive cost for real models, whose depths are of the order of hundreds (see Table~\ref{tab:real_CNNs_description}). For example, there are more than $3 \cdot 10^9$ possibilities for {\sc Resnet101} with $s=6$ fragments (the minimum needed to avoid host memory), since it has depth $d=209$.

\input{mem_usage_profiling_synthetic_4_seg}

\begin{figure}[t]
    \centering
    \includegraphics[width=0.85\textwidth]{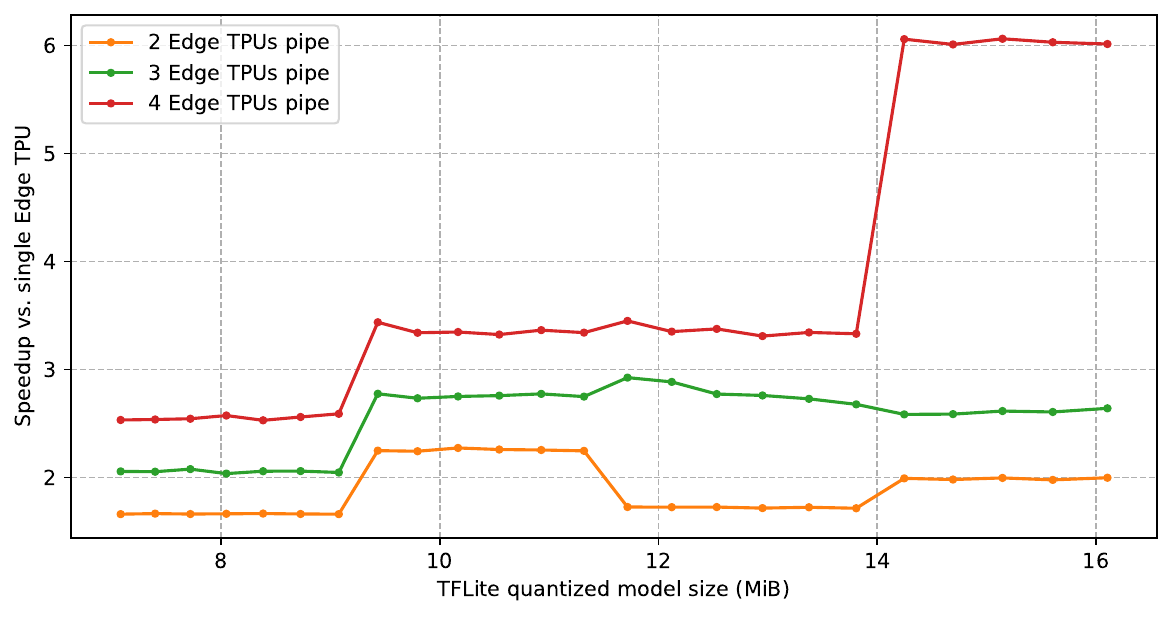}
    \caption{Speedup of synthetic models segmented using \SEGMPROF into $2$, $3$ and $4$ TPUs run on a $15$-input batch, versus execution on a single TPU.}
    \label{fig:speedup_profiling_vs_1_TPU}
\end{figure}

In Table~\ref{tab:mem_usage_synthetic_4_seg} we saw that the compiler underutilized the memory of the first TPU when segmenting our synthetic models into $4$ fragments, and overutilized the memory of the fourth one, so it was necessary to use host memory in some cases. This contrasts with the memory usage of the profiling-based segmentation shown in Table~\ref{tab:mem_usage_profiling}. In all cases we see that the fragments are very similar in size and, in fact, no model requires using host memory. Those partitions are chosen because they obtain the best performance (in particular better than the compiler ones), which confirms that it is preferable to avoid the use of host memory and that it is better to balance the size even if it does not reduce/avoid the use of host memory (some compiler partitions did not use it). More size balancing presumably implies more workload balancing.

Figure~\ref{fig:speedup_profiling_vs_1_TPU} shows the speedup obtained in the inference of these models with 
\SEGMPROF for $2$, $3$ and $4$ TPUs, versus execution on a single TPU. Contrary to \SEGMCOMP (see Figure~\ref{fig:speedup_comp_seg_vs_1_TPU}), the speedups are quite close to the number of TPUs used and, 
indeed, notably higher for the larger models: with $4$ TPUs, a $6 \times$ is obtained because a significant 
host usage is completely avoided. 
As we already stated, speeding up proportionally to the number of devices is not enough because, in addition to 
parallelizing the operations, we are reducing communications with the host; however, the $6 \times$ improvements with $4$ 
devices are really interesting.

%% file: mem_usage_synthetic_4_seg.tex
\begin{table*}[t]
\centering

\setlength{\tabcolsep}{5pt}

\caption{Memory usage of synthetic models split into $4$ parts using \SEGMCOMP.}
\label{tab:mem_usage_synthetic_4_seg}

\setlength{\belowrulesep}{0pt}
    \begin{tabular}{c|cccccccc|}\cmidrule{2-9} 
                           & \multicolumn{8}{c|}{Host and device memory usage of each TPU (MiB)} \\ \hline
    \rowcolor[HTML]{EFEFEF} 
    \multicolumn{1}{|c|}{\cellcolor[HTML]{EFEFEF}\begin{tabular}[c]{@{}c@{}}Model size\\ (MiB)\end{tabular}} &
      Dev. 1 &
      Dev. 2 &
      Dev. 3 &
      Dev. 4 &
      Host 1 &
      Host 2 &
      Host 3 &
      Host 4 \\ \hline
    \multicolumn{1}{|c|}{8.04} & 0.021  & 2.00  & 2.00  & 4.01 & 0 & 0 & 0 & 0       \\
    \multicolumn{1}{|c|}{9.08} & 0.022  & 2.26  & 2.26  & 4.53 & 0 & 0 & 0 & 0       \\
    \multicolumn{1}{|c|}{10.17} & 0.024  & 2.54  & 2.54  & 5.07 & 0 & 0 & 0 & 0       \\
    \multicolumn{1}{|c|}{11.31} & 0.025  & 2.82  & 2.82  & 5.64 & 0 & 0 & 0 & 0       \\
    \multicolumn{1}{|c|}{12.53} & 0.026  & 3.13  & 3.13  & 3.13 & 0 & 0 & 0 & 3.13    \\
    \multicolumn{1}{|c|}{13.81} & 0.027  & 3.44  & 3.44  & 3.44 & 0 & 0 & 0 & 3.44    \\
    \multicolumn{1}{|c|}{15.14} & 0.029  & 3.78  & 3.78  & 3.78 & 0 & 0 & 0 & 3.78      \\
    \multicolumn{1}{|c|}{16.60} & 0.030  & 4.08  & 4.08  & 4.08 & 0 & 0 & 0 & 4.08      \\ \hline
    \end{tabular}
\end{table*}

%% file: real_models_compiler_segmentation.tex
\setlength{\tabcolsep}{4pt}
\begin{table}[t]
\centering
\caption{Compiler segmentation (\SEGMCOMP) results on real models, compared to a single TPU in terms of host memory usage and inference time.}
    \label{tab:real_models_comp_segm}

    \setlength{\belowrulesep}{0pt}
    \begin{tabular}{lc|cc|c||cc||c|}\cmidrule{3-5}\cmidrule{6-8}
     &  & \multicolumn{3}{c||}{Host memory (MiB)} & \multicolumn{2}{c||}{Inference time (ms)} & Speedup \\ \hline
    \rowcolor[HTML]{EFEFEF} 
    \multicolumn{1}{|c|}{\cellcolor[HTML]{EFEFEF}Model} & \begin{tabular}[c]{@{}c@{}}Num.\\ TPUs\footnotemark[1]\end{tabular} & \multicolumn{1}{c|}{\cellcolor[HTML]{EFEFEF}1 TPU} & \begin{tabular}[c]{@{}c@{}}{\sc Segm\_comp}\end{tabular} & \begin{tabular}[c]{@{}c@{}}$\Delta_s$ \\ (MiB)\footnotemark[2]\end{tabular} & \multicolumn{1}{c|}{\cellcolor[HTML]{EFEFEF}1 TPU} & \begin{tabular}[c]{@{}c@{}}{\sc Segm\_comp}\end{tabular} & \multicolumn{1}{c|}{\cellcolor[HTML]{EFEFEF}\begin{tabular}[c]{@{}c@{}}{\sc Segm\_comp}\\ vs. 1 TPU\end{tabular}} \\ \hline
    \multicolumn{1}{|l|}{Xception} & 4 & \multicolumn{1}{c|}{17.72} & 0 & 2.15 & \multicolumn{1}{c|}{60.11} & 16.60 & \multicolumn{1}{c|}{3.62$\times$ (0.90$\times$)} \\
    \multicolumn{1}{|l|}{ResNet50} & 4 & \multicolumn{1}{c|}{17.54} & 0 & 1.86 & \multicolumn{1}{c|}{29.69} & 7.60 & \multicolumn{1}{c|}{3.91$\times$ (0.97$\times$)} \\
    \multicolumn{1}{|l|}{ResNet50V2} & 4 & \multicolumn{1}{c|}{17.96} & 0 & 1.88 & \multicolumn{1}{c|}{30.94} & 8.15 & \multicolumn{1}{c|}{3.80$\times$ (0.95$\times$)} \\
    \multicolumn{1}{|l|}{ResNet101} & 6 & \multicolumn{1}{c|}{35.90} & \cellcolor[HTML]{FAB5BE}2.03 & 2.34 & \multicolumn{1}{c|}{44.73} & 11.58 & \multicolumn{1}{c|}{3.86$\times$ (0.64$\times$)} \\
    \multicolumn{1}{|l|}{ResNet101V2} & 6 & \multicolumn{1}{c|}{36.83} & \cellcolor[HTML]{FAB5BE}2.07 & 2.31 & \multicolumn{1}{c|}{54.94} & 11.33 & \multicolumn{1}{c|}{4.85$\times$ (0.80$\times$)} \\
    \multicolumn{1}{|l|}{ResNet152} & 8 & \multicolumn{1}{c|}{51.04} & \cellcolor[HTML]{FAB5BE}2.13 & 2.21 & \multicolumn{1}{c|}{68.94} & 12.62 & \multicolumn{1}{c|}{5.46$\times$ (0.68$\times$)} \\
    \multicolumn{1}{|l|}{ResNet152V2} & 8 & \multicolumn{1}{c|}{52.42} & \cellcolor[HTML]{FAB5BE}2.13 & 2.21 & \multicolumn{1}{c|}{72.84} & 12.87 & \multicolumn{1}{c|}{5.66$\times$ (0.70$\times$)} \\
    \multicolumn{1}{|l|}{InceptionV3} & 4 & \multicolumn{1}{c|}{17.97} & \cellcolor[HTML]{FFD7D5}0.56 & 2.04 & \multicolumn{1}{c|}{36.96} & 11.24 & \multicolumn{1}{c|}{3.29$\times$ (0.82$\times$)} \\
    \multicolumn{1}{|l|}{InceptionV4} & 7 & \multicolumn{1}{c|}{36.30} & \cellcolor[HTML]{FFD7D5}0.95 & 2.12 & \multicolumn{1}{c|}{82.73} & 13.94 & \multicolumn{1}{c|}{5.93$\times$ (0.84$\times$)} \\
    \multicolumn{1}{|l|}{Inc.ResNetV2} & 8 & \multicolumn{1}{c|}{49.61} & \cellcolor[HTML]{FAB5BE}3.27 & 2.85 & \multicolumn{1}{c|}{86.87} & 21.55 & \multicolumn{1}{c|}{4.03$\times$ (0.50$\times$)} \\
    \multicolumn{1}{|l|}{DenseNet121} & 2 & \multicolumn{1}{c|}{2.98} & 0 & 1.70 & \multicolumn{1}{c|}{14.88} & 8.52 & \multicolumn{1}{c|}{1.75$\times$ (0.87$\times$)} \\
    \multicolumn{1}{|l|}{DenseNet169} & 3 & \multicolumn{1}{c|}{8.59} & 0 & 1.82 & \multicolumn{1}{c|}{30.94} & 12.97 & \multicolumn{1}{c|}{2.39$\times$ (0.79$\times$)} \\
    \multicolumn{1}{|l|}{DenseNet201} & 4 & \multicolumn{1}{c|}{15.17} & 0 & 1.88 & \multicolumn{1}{c|}{50.12} & 14.11 & \multicolumn{1}{c|}{3.55$\times$ (0.88$\times$)} \\
    \multicolumn{1}{|l|}{Eff.NetLiteB3} & 2 & \multicolumn{1}{c|}{2.18} & 0 & \cellcolor[HTML]{D8F8B8}0.23 & \multicolumn{1}{c|}{10.31} & 3.96 & \multicolumn{1}{c|}{2.60$\times$ (1.30$\times$)} \\
    \multicolumn{1}{|l|}{Eff.NetLiteB3} & 3 & \multicolumn{1}{c|}{7.93} & 0 & \cellcolor[HTML]{D8F8B8}0.41 & \multicolumn{1}{c|}{38.17} & 10.99 & \multicolumn{1}{c|}{3.47$\times$ (1.15$\times$)} \\ \hline
    \end{tabular}
\footnotetext[1]{Number of TPUs used to evaluate the compiler segmentation (i.e. the number of segments). We use the minimum number of TPUs that would ideally avoid host memory usage.}
\footnotetext[2]{Difference between the size of the largest and smallest segment produced by the compiler.}
\end{table}
\setlength{\tabcolsep}{6pt}

%% file: mem_usage_profiling_synthetic_4_seg.tex
\begin{table}[t]
\centering
\caption{Memory usage of synthetic models split into $4$ parts with \SEGMPROF.}
\label{tab:mem_usage_profiling}

\setlength{\belowrulesep}{0pt}
    \begin{tabular}{c|cccccccc|}\cmidrule{2-9}
                           & \multicolumn{8}{c|}{Host and device memory usage of each TPU (MiB)} \\ \hline
    \rowcolor[HTML]{EFEFEF} 
    \multicolumn{1}{|c|}{\cellcolor[HTML]{EFEFEF}\begin{tabular}[c]{@{}c@{}}Model size\\ (MiB)\end{tabular}} &
      Dev. 1 &
      Dev. 2 &
      Dev. 3 &
      Dev. 4 &
      Host 1 &
      Host 2 &
      Host 3 &
      Host 4 \\ \hline
    \multicolumn{1}{|c|}{8.04} & 2.02  & 2.00  & 2.00  & 2.01 & 0 & 0 & 0 & 0       \\
    \multicolumn{1}{|c|}{9.08} & 2.28  & 2.26  & 2.26  & 2.26 & 0 & 0 & 0 & 0       \\
    \multicolumn{1}{|c|}{10.17} & 2.56  & 2.54  & 2.54  & 2.54 & 0 & 0 & 0 & 0       \\
    \multicolumn{1}{|c|}{11.31} & 2.85  & 2.82  & 2.82  & 2.82 & 0 & 0 & 0 & 0       \\
    \multicolumn{1}{|c|}{12.53} & 3.13  & 3.13  & 3.13  & 3.13 & 0 & 0 & 0 & 0    \\
    \multicolumn{1}{|c|}{13.81} & 3.47  & 3.44  & 3.44  & 3.44 & 0 & 0 & 0 & 0    \\
    \multicolumn{1}{|c|}{15.14} & 3.81  & 3.78  & 3.78  & 3.78 & 0 & 0 & 0 & 0      \\
    \multicolumn{1}{|c|}{16.60} & 4.11  & 4.08  & 4.08  & 4.08 & 0 & 0 & 0 & 0      \\ \hline
    \end{tabular}
\end{table}

%% file: s6-balanced_segmentation.tex
\section{Balanced model segmentation ({\sc Segm\_Balanced})\label{sec:balanced_segm}}

As mentioned above, optimizing segmentation by exhaustive analysis of all possibilities is not a 
feasible option for most real-world models. 
To tackle this problem, a new segmentation scheme to solve the unbalance problems present in the compiler with an affordable computational cost and targeting real models is needed. The following sections are devoted to explain in detail our proposed segmentation process and evaluate its results. We will refer to this balanced segmentation
strategy as \SEGMBALANCED.

\subsection{Segmentation process}

\SEGMBALANCED consists of three consecutive steps: 
{\em (i)} depth-based layer location (described in Section~\ref{sec:depth-based_layer_location}); 
{\em (ii)} balancing model segmentation (Section~\ref{sec:balancing_model_segmentation}); and 
{\em (iii)} refining the segmentation to reduce the host memory usage (Section~\ref{sec:refining_segmentation}).

\subsubsection{Depth-based layer location}\label{sec:depth-based_layer_location}

\begin{figure}[t]
    \centering
    \includegraphics[width=0.7\textwidth]{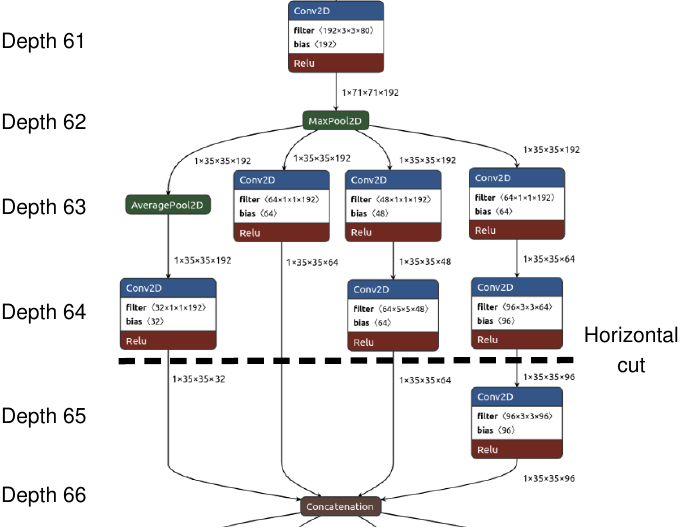}
    \caption{{\sc InceptionV3} architecture block with four open paths and illustration of a possible horizontal cut.}
    \label{fig:inceptionV3_multipath_levels}
\end{figure}

For the synthetic models, all networks were composed by a chain of layers, making the segmentation decision 
a rather simple process. 
Indeed, the separation points have to be between the last layer of a segment and the first layer of the next one. 
However, not all real models feature a ``chain-based'' layered structure, 
having the output of some layers not directly connected to only the next layer, 
but to (possibly) more than one layer. 
For example, Figure~\ref{fig:inceptionV3_multipath_levels} shows a part of the network {\sc InceptionV3} model, 
where the output of a {\em MaxPool2D} layer is connected to the input of four other layers, giving rise to four open paths. 
This implies that, for an optimal segmentation of the model, it needs to be separated by all open paths to form 
correct disjoint segments.

When there are several paths open, it is possible to segment the network by cutting the connection between nodes at different depths. For example, we could cut the leftmost path of Figure~\ref{fig:inceptionV3_multipath_levels} between the \textit{MaxPool2D} layer and the \textit{AveragePool2D} layer (just after depth $62$), and cut the other 3 paths between the last \textit{Conv2D} layer and the Concatenation layer (just after depths $63$, $64$ and $65$ respectively). However, considering all these cuts increases significantly the space of possible fragmentations and therefore, the complexity of the algorithm. 
Indeed, performing the split process using only horizontal cuts (i.e., separations of all paths at the same depth), allows enough flexibility to distribute the workload and size of the model in many fragments, offering a good trade-off between optimality and performance, as proved later in the experimental section. Figure~\ref{fig:inceptionV3_multipath_levels} shows a horizontal cut as considered by \SEGMBALANCED.

In order to perform these cuts, an identification of the depth of each layer in the model is needed. 
To calculate the depth of the layers, we treat the models as directed acyclic  graphs (DAGs) because the networks are feed-forward. In this type of graphs we can calculate the topological order of the nodes and use it to find the maximum distance of each one from the input, i.e. its depth~\cite{sedgewick2011}.

\subsubsection{Balancing model segmentation}\label{sec:balancing_model_segmentation}

The main problem of \SEGMCOMP is that the produced segments present very unequal sizes: some TPUs receive very large fragments that need to be partially stored on the host, while others are under-utilized with small fragments. In this sense, a good approach to improve this segmentation is to choose the partition that minimizes the size of the largest segment.

The size occupied by a model depth level is proportional to the sum of the number of parameters of its layers (in fact, it is the same in the 8-bit quantized model as each parameter occupies one byte).
In this way, we can represent the sizes of the model by depth as an array of elements $P = [P_0, ..., P_{d-1}]$, where $P_i$ is the number of parameters at depth $i$ and $d$ is the total depth of the model. Now the problem has transformed into splitting the $P$ array by minimizing the maximum sum of the subarrays.

That problem can be optimally solved with a binary search over the possible upper bounds for the sum of the subarrays (values up to $\sum_{i=0}^{d-1} P_i$), as shown in Algorithm~\ref{alg:balancedSplit}. The \texttt{balancedSplit} function implements a binary search that calls at each step to the greedy method \texttt{splitCheck} to verify if the array $P$ can be segmented into $s$ parts with at most $bound$ parameters in each. This method traverses the array assigning values to a fragment as long as their sum is less than $bound$, and starting with a new fragment when exceeding it (lines 19-24). If the final amount of fragments needed does not exceed the required $s$, then $bound$ is a sufficient upper bound and smaller values are searched (lines 8-10). If it is not, the binary search moves to larger values (line 12). The greedy method also returns the split positions found, which are updated in the binary search when a lower upper bound is found. Thus, after the search we will know the depth levels to segment with the minimum upper bound.

\input{balanced_split}

Note that the time complexity of this phase is $\mathcal{O}\big(d \cdot log(\sum_{i=0}^{d-1} P_i)\big)$, which is affordable for the models. The depths of our models are in the order of hundreds and the number of parameters is in the tens of millions of parameters (see Table~\ref{tab:real_CNNs_description}), so its logarithm is not too large. For example, in {\sc ResNet101} with $d=209$ and $44.7$ million parameters, the number of operations will be on the order of $209 \cdot log(44.7 \cdot 10^6) \simeq 5311$.

\subsubsection{Refining segmentation to reduce host memory usage}\label{sec:refining_segmentation}

Although the previous phase minimizes the size of the largest fragment and produces segments of balanced size (calculated in terms of the number of parameters), there is some variation in the amount of memory needed when compiling for the Edge TPU. 
Indeed, additional memory space is needed to save the inputs and network activations. In addition, we have observed some extra memory usage that we attribute to alignment and padding issues. These aspects are difficult to estimate in our previous segmentation decision. However, as our current segmentation is very close to a balanced memory usage after compiling, we can simply use the compiler report as feedback to slightly readjust it.

\begin{figure}[t]
    \centering
    \includegraphics[width=1\textwidth]{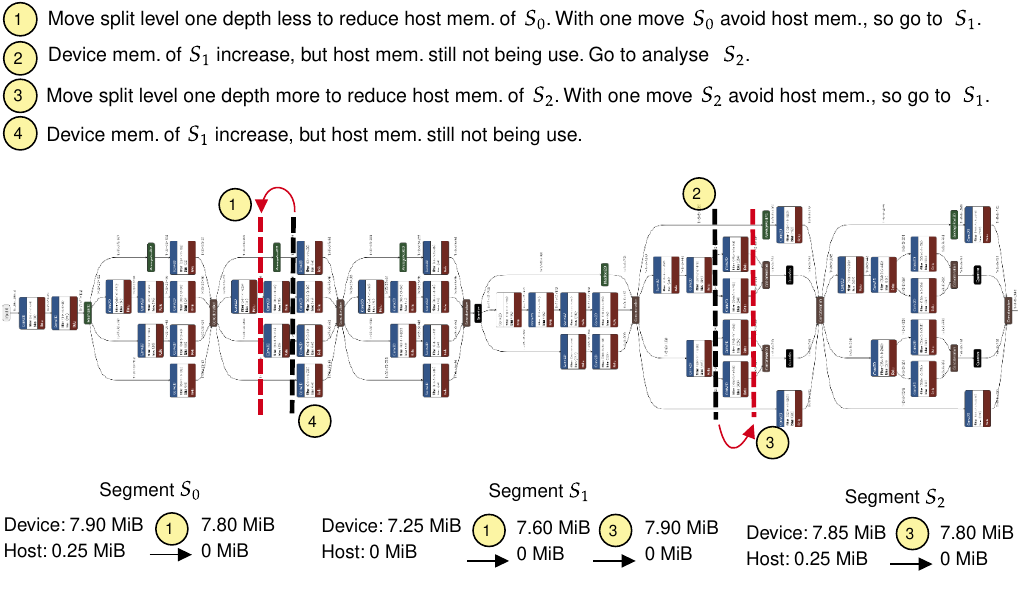}
    \caption{Example diagram of a segmentation refinement. Top: Description of iterations. Middle: Split point analyzed in each iteration (indicated with numbers) and its movement (indicated by arrows). Bottom: Variation of device and host memory usages.}
    \label{fig:refining_segmentation}
\end{figure}

Our goal is to avoid the use of external memory in all TPUs, because it is the main bottleneck. To achieve it, we move the split point of each segment $S_i$ with the next segment $S_{i+1}$ to one less depth level in case $S_i$ uses host memory when compiled (see example in Figure~\ref{fig:refining_segmentation}). In this way, the size of $S_i$ is reduced and the size of $S_{i+1}$ is increased by the same amount (i.e., shifting layers from one to the other). This process is repeated until segment $S_i$ does not use host memory and then starts with the next segment, $S_{i+1}$. Note that traversing from the first segment to the last one does not work correctly if the last one must be reduced by moving some layers to previous segments (in fact, the process tends to move layers to the last segment). Because of this, in these situations, the same process is repeated from the last segment to the first by moving the splitting point to deeper levels (exemplified in Figure~\ref{fig:refining_segmentation}).


Although each compilation may even take a few seconds, the refinement process is not very expensive as the segmentation of~\ref{sec:balancing_model_segmentation} is good enough, and the split points move very rarely. However, the process can be optimized by moving the split points several positions at a time depending on the host memory usage and the size of depths prior the split point. Thus, when a split point has to be moved $x$ positions, a single iteration (with a single compilation) is done instead of $x$ iterations (with $x$ compilations).

\subsection{Performance evaluation}

Finally, we evaluate the performance of both model families with the proposed segmentation scheme and compare it with the other strategies. Again, the evaluation has been performed on a $15$-input batch, with $2$, $3$ and $4$ TPUs for the synthetic models, and with the minimum number of TPUs that would ideally allow avoiding host memory usage for the real models.

For the synthetic models, our segmentation always obtains the best performing partition, found by brute force by the profiling-based strategy we saw in Section~\ref{sec:profiling}. The results are exactly the same as the ones described on that section. In fact, no partition refinement was needed on any model of Section~\ref{sec:refining_segmentation}, since the partition achieved by the balanced parameter partitioning of \ref{sec:balancing_model_segmentation} already avoided the use of host memory. Note that these models consist of few very large layers, so there are few possible partitions and it is natural that our layer-size driven scheme obtains the optimal solution; however, this improves the compiler's approach which made much worse partitions.

On the other hand, Table~\ref{tab:real_model_segm_results} shows the results obtained by our segmentation proposal when applied to real models, compared to the execution on a single TPU and the compiler segmentation. The results reveal the following highlights:

\begin{itemize}

    \item \SEGMBALANCED improves the performance of the compiler strategy for all models. It improves especially (marked in green in Table~\ref{tab:real_model_segm_results}) when the compiler uses host memory (see red cells in Table~\ref{tab:real_models_comp_segm}): $2.6\times $ when using more than $3$ MiB, around $2\times$ when using about $2$ MiB, and around $1.6\times$ when using less than $1$ MiB. In other models where the compiler does not use host memory, inference is about $1.4 \times$ faster, presumably by more efficient pipeline usage. The exceptions are the \textsc{EfficientNetLite} models because the compiler splits them into fairly balanced sizes, as we saw in~\ref{sec:compiler_segmentation}.
    
    \item Even though it is not shown in the table, \SEGMBALANCED manages to avoid the use of host memory in all models, including the seven cases where compiler segmentation still needed to use it (marked in red in Table~\ref{tab:real_models_comp_segm}).

    \item The speedup of \SEGMBALANCED approach versus a single TPU is higher than the number of TPUs used in all models. Our strategy is very profitable in terms of performance for all of them, even taking into account the extra hardware required.
\end{itemize}

\input{real_models_segm_results}

\begin{figure}[t!]
    \centering
    \includegraphics[width=0.8\textwidth]{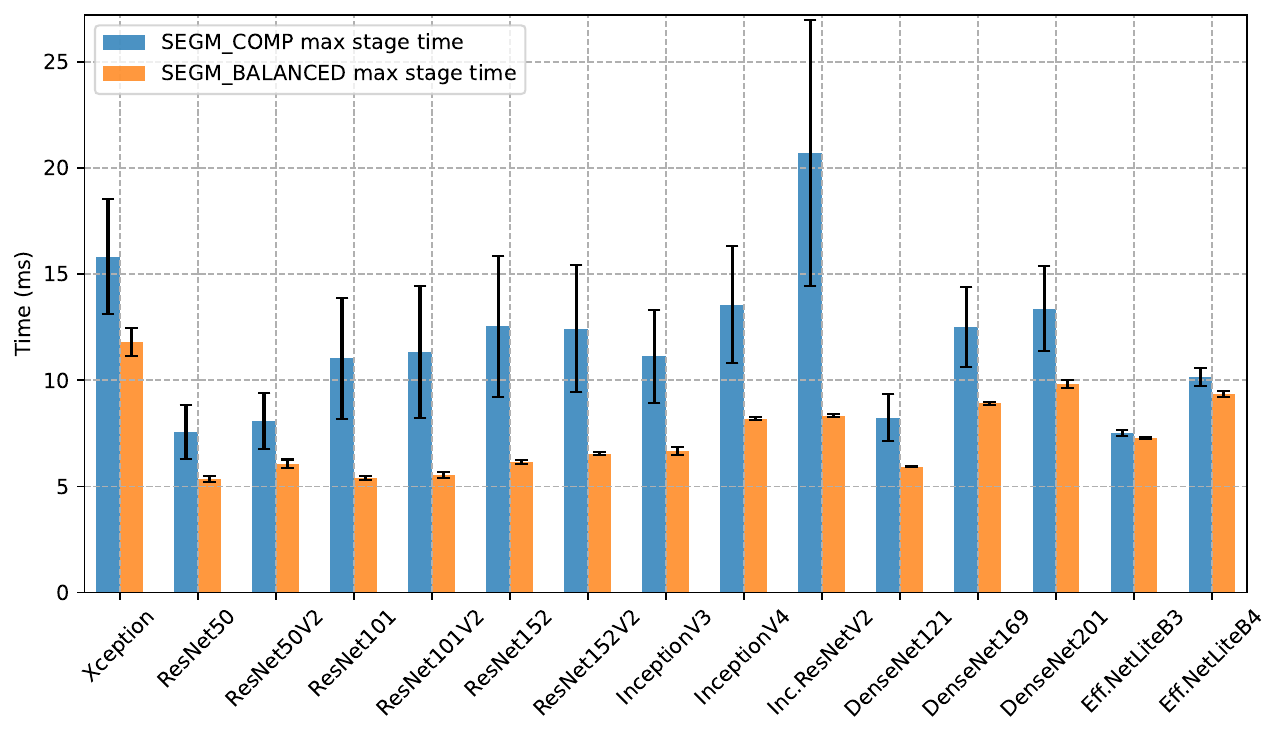}
    \caption{Bars with the time of the slowest stage of the real models with both segmentations, and segments with the difference with respect to the average time of the stages.}
    \label{fig:times_max_stages_and_deviation}
    \vspace{-10pt}
\end{figure}

We should also mention that \SEGMBALANCED only had to perform the refinement step (Section~\ref{sec:refining_segmentation}) for 5 of the 15 real models (in the rest the use of host memory was avoided thanks to the algorithm described in~\ref{sec:balancing_model_segmentation}). Although that step is relatively expensive as it requires several compilations of the different fragments, the segmentation of the models that ran this step took less than a minute on an i9-9900K CPU (versus less than a second for those that did not). Considering that we only have to segment the model once, these times are quite reasonable. Moreover, the process was performed without the possible optimizations commented at the end of Section~\ref{sec:refining_segmentation} to speed up this step.

To clearly illustrate why \SEGMBALANCED improves when compared with \SEGMCOMP even when host memory usage is not reduced, Figure~\ref{fig:times_max_stages_and_deviation} shows the times of the slowest stages with each strategy and their deviation from the mean time. It is clearly seen that the slowest stages with the \SEGMCOMP strategy take more time than the slowest stages with \SEGMBALANCED, resulting in worse overall performance since the slowest stage limits the performance of the pipeline. A significant difference between the maximum and average time is seen in the compiler strategy, especially in models using host memory due to slow inter-device communications, which are solved by achieving an almost perfect partitioning with our scheme. It is also observed that the compiler makes a very good partitioning of the \textsc{EfficientNetLite3}, since the time of the slowest stage is practically the average time. In such a case there is almost no margin for improvement.

%% file: balanced_split.tex
\algtext*{EndWhile}
\algtext*{EndIf}
\algtext*{EndFunction}
\algtext*{EndFor}
\begin{algorithm}
\begin{algorithmic}[1]
\caption{Balancing model segmentation pseudocode.}
\label{alg:balancedSplit}
\Function{balancedSplit}{$P[\,]$, $s$}
\State $minSearch := max(P)$\Comment{An upper bound must exceed all elements}
\State $maxSearch := sum(P)$\Comment{The array sum is an obvious upper bound}
\State $bestSplit := nothing$
\While{$minSearch \leq maxSearch$} \Comment{Binary search loop}
    \State $posBound := (minSearch + maxSearch)\,/\,2$
    \State $check,\,splitPos :=  splitCheck(P,\,bound,\,s)$ \Comment{Call for bound check}
    \If{$check$} \Comment{If $bound$ is an upper bound}
        \State $bestSplit := splitPos$ \Comment{Save the cut-off points}
        \State $maxSearch := bound - 1$ \Comment{Search for smaller upper bounds}
    \Else \Comment{If $bound$ is not an upper bound}
        \State $minSearch := bound + 1$ \Comment{Search for greater upper bounds}
    \EndIf
\EndWhile
\State \textbf{return} $bestSplit$
\EndFunction\\

\Function{splitCheck}{$P[\,]$, $bound$, $s$}
  \State $minSegms := 0$
  \State $paramsSum := 0$
  \State $splitPos := [\,]$
  \For{$i:= 0$ \textbf{to} $length(P) - 1$} \Comment{For each depth level}
        \State $paramsSum := paramsSum + P[i]$ \Comment{Accumulate its params}
        \If{$paramsSum > bound$} \Comment{If the upper bound is exceeded}
            \State  $append(splitPos, i-1)$ \Comment{Add cut just before this depth}
            \State  $minSegms := minSegms + 1$ \Comment{Count one more segment}
            \State $paramsSum := P[i]$ \Comment{Init params sum of next segment}
        \EndIf
  \EndFor
  \State $minSegms := minSegms + 1$ \Comment{Count the last segment}
  \State $check := minSegms \leq s$ \Comment{Check if $s$ fragments are enough}
  \State \textbf{return} $check,\,splitPos$\Comment{Return check and cut-off points}
\EndFunction

\end{algorithmic}
\end{algorithm}

%% file: real_models_segm_results.tex
\setlength{\tabcolsep}{3.75pt}

\begin{table}[t]
\setlength{\belowrulesep}{0pt}
\centering
\caption{Multi-TPU inference performance results for real models with {\sc Segm\_balanced} and {\sc Segm\_comp} versus a single TPU execution.}
\label{tab:real_model_segm_results}

\begin{tabular}{lc|c|c|c||c|c|}\cmidrule{3-7}
   &
   &
  \multicolumn{3}{c||}{Inference time (ms)} &
  \multicolumn{2}{c|}{Speedup} \\ \hline
\rowcolor[HTML]{EFEFEF} 
\multicolumn{1}{|c|}{\cellcolor[HTML]{EFEFEF}Model} &
  \begin{tabular}[c]{@{}c@{}}Num.\\ TPUs\end{tabular} &
  1 TPU &
  \begin{tabular}[c]{@{}c@{}}{\sc Segm\_comp}\end{tabular} &
  \begin{tabular}[c]{@{}c@{}}{\sc Segm\_balanced}\end{tabular} &
  \begin{tabular}[c]{@{}c@{}}{\sc Segm\_balanced}\\ vs {\sc Segm\_comp}.\end{tabular} &
  \begin{tabular}[c]{@{}c@{}}{\sc Segm\_balanced}\\ vs 1 TPU\end{tabular} \\ \hline
\multicolumn{1}{|l|}{Xception} &
  4 &
  60.11 &
  16.60 &
  12.64 &
  $1.31 \times$&
  $4.76 \times$ ($1.19 \times$)\\
\multicolumn{1}{|l|}{ResNet50} &
  4 &
  29.69 &
  7.60 &
  5.28 &
  $1.44 \times$&
  $5.62 \times$ ($1.40 \times$)\\
\multicolumn{1}{|l|}{ResNet50V2} &
  4 &
  30.94 &
  8.15 &
  6.13 &
  $1.33 \times$&
  $5.05 \times$ ($1.26 \times$)\\
\multicolumn{1}{|l|}{ResNet101} &
  6 &
  44.73 &
  11.58 &
  5.59 &
  \cellcolor[HTML]{B8E986}$2.07 \times$&
  $8.00 \times$ ($1.33 \times$)\\
\multicolumn{1}{|l|}{ResNet101V2} &
  6 &
  54.94 &
  11.33 &
  5.52 &
  \cellcolor[HTML]{B8E986}$2.05 \times$&
  $8.43 \times$ ($1.40 \times$)\\
\multicolumn{1}{|l|}{ResNet152} &
  8 &
  68.94 &
  12.62 &
  6.30 &
  \cellcolor[HTML]{B8E986}$2.00 \times$&
  $10.94 \times$ ($1.36 \times$)\\
\multicolumn{1}{|l|}{ResNet152V2} &
  8 &
  72.84 &
  12.87 &
  6.63 &
  \cellcolor[HTML]{B8E986}$1.94 \times$&
  $10.99 \times$ ($1.37 \times$)\\
\multicolumn{1}{|l|}{InceptionV3} &
  4 &
  36.96 &
  11.24 &
  6.72 &
  \cellcolor[HTML]{D8F8B8}$1.67 \times$&
  $5.50 \times$ ($1.37 \times$)\\
\multicolumn{1}{|l|}{InceptionV4} &
  7 &
  82.73 &
  13.94 &
  8.69 &
  \cellcolor[HTML]{D8F8B8}$1.60 \times$&
  $9.52 \times$ ($1.36 \times$)\\
\multicolumn{1}{|l|}{Inc.ResNetV2} &
  8 &
  86.87 &
  21.55 &
  8.28 &
  \cellcolor[HTML]{B8E986}$2.60 \times$&
  $10.49 \times$ ($1.31 \times$)\\
\multicolumn{1}{|l|}{DenseNet121} &
  2 &
  14.88 &
  8.52 &
  6.05 &
  $1.41 \times$&
  $2.46 \times$ ($1.23 \times$)\\
\multicolumn{1}{|l|}{DenseNet169} &
  3 &
  30.94 &
  12.97 &
  8.96 &
  $1.45 \times$&
  $3.45 \times$ ($1.15 \times$)\\
\multicolumn{1}{|l|}{DenseNet201} &
  4 &
  50.12 &
  14.11 &
  10.13 &
  $1.39 \times$&
  $4.95 \times$ ($1.23 \times$)\\
\multicolumn{1}{|l|}{Eff.NetLiteB3} &
  2 &
  10.31 &
  3.96 &
  3.88 &
  $1.02 \times$&
  $2.66 \times$ ($1.33 \times$)\\
\multicolumn{1}{|l|}{Eff.NetLiteB4} &
  3 &
  38.17 &
  10.99 &
  10.68 &
  $1.03 \times$&
  $3.57 \times$ ($1.19 \times$)\\ \hline
\end{tabular}
\end{table}
\setlength{\tabcolsep}{6pt}

%% file: s7-conclusions.tex
\section{Conclusions}\label{sec:conclusions}

In this paper, we have addressed the main limitations of the Edge TPU by Google for 
inference on real-world CNN models. Specifically, after evaluating and characterizing the
performance of both synthetic models and state-of-the-art CNNs, we have concluded that
the scarce amount of on-chip memory per Edge TPU becomes the main bottleneck when using it
for large CNNs. 

Our profile-based balanced segmentation strategy aims, at the same time, at leveraging multiple
Edge TPUs to improve inference performance, but also to alleviate the memory transfer bottleneck
by spreading models across on-chip device memory, while at the same time improving workload balance
across TPUs. 

Together, these proposals overperform both the theoretical peak performance on multiple TPUs considering
the baseline performance on a single one, but also that attained by the vendor's compiler, yielding better
performance even for models that fit across the available on-chip memory. The attained performance acceleration
(up to $2.60\times$ compared with a compiler-based pipelined segmentation), and super-linear compared with
a single TPU setup, validate our approach towards multi-TPU exploitation for inference on CNNs.

As a future work, we plan to extend our study to other CNNs, including networks with more complex topologies, 
and to compare both the performance and energy efficiency of multi-TPU setups compared with alternative architectures
(mainly single and multi-GPU setups).

%% file: main.bbl
\begin{thebibliography}{45}
\providecommand{\natexlab}[1]{#1}
\providecommand{\url}[1]{{#1}}
\providecommand{\urlprefix}{URL }
\providecommand{\doi}[1]{\url{https://doi.org/#1}}
\providecommand{\eprint}[2][]{\url{#2}}
 \bibcommenthead

\bibitem[{Alshehri and Muhammad(2021)}]{Alshehri2021ACS}
Alshehri F, Muhammad G (2021) {A Comprehensive Survey of the Internet of Things (IoT) and AI-Based Smart Healthcare}. IEEE Access 9:3660--3678. \doi{10.1109/ACCESS.2020.3047960}

\bibitem[{Antonini et~al(2019)Antonini, Vu, Min, Montanari, Mathur, and Kawsar}]{Antonini2019}
Antonini M, Vu TH, Min C, et~al (2019) Resource characterisation of personal-scale sensing models on edge accelerators. In: Int. Workshop on Challenges in Artificial Intelligence and Machine Learning for Internet of Things, p 49–55, \doi{10.1145/3363347.3363363}

\bibitem[{{ASUS}(2023)}]{asus_datasheet}
{ASUS} (2023) {ASUS CRL-G18U-P3DF Datasheet}. \urlprefix\url{https://dlcdnets.asus.com/pub/ASUS/mb/AIOT/AI_Accelerator/AI_Accelerator_Card_Spec_Sheet.pdf?model=CRL-G18U-P3DF}

\bibitem[{Cass(2019)}]{Cass2019}
Cass S (2019) {Taking AI to the edge: Google's TPU now comes in a maker-friendly package}. IEEE Spectrum 56(5):16--17. \doi{10.1109/MSPEC.2019.8701189}

\bibitem[{Du et~al(2021)Du, Zhu, Shen, Du, Lu, Xiao, and Liao}]{Du2021}
Du J, Zhu X, Shen M, et~al (2021) Model parallelism optimization for distributed inference via decoupled cnn structure. IEEE Transactions on Parallel and Distributed Systems 32(7):1665--1676. \doi{10.1109/TPDS.2020.3041474}

\bibitem[{{Google Coral}(2019)}]{coral_tech}
{Google Coral} (2019) {M.2 Accelerator A+E key datasheet}. \urlprefix\url{https://coral.ai/docs/m2/datasheet/}

\bibitem[{Guo et~al(2019)Guo, Liu, Wang, Yao, Han, Li, Lu, and Hu}]{Guo2019}
Guo J, Liu W, Wang W, et~al (2019) Accudnn: A gpu memory efficient accelerator for training ultra-deep neural networks. In: 2019 IEEE 37th International Conference on Computer Design (ICCD), pp 65--72, \doi{10.1109/ICCD46524.2019.00017}

\bibitem[{He et~al(2020)He, Guo, Guo, Qiu, and Qi}]{He2020}
He W, Guo S, Guo S, et~al (2020) Joint dnn partition deployment and resource allocation for delay-sensitive deep learning inference in iot. IEEE Internet of Things Journal 7(10):9241--9254. \doi{10.1109/JIOT.2020.2981338}

\bibitem[{Hu et~al(2019)Hu, Bao, Wang, and Liu}]{Hu2019}
Hu C, Bao W, Wang D, et~al (2019) Dynamic adaptive dnn surgery for inference acceleration on the edge. In: IEEE INFOCOM 2019-IEEE Conference on Computer Communications, IEEE, pp 1423--1431, \doi{10.1109/INFOCOM.2019.8737614}

\bibitem[{Huang et~al(2019)Huang, Cheng, Bapna, Firat, Chen, Chen, Lee, Ngiam, Le, Wu, and Chen}]{Huang2019}
Huang Y, Cheng Y, Bapna A, et~al (2019) Gpipe: Efficient training of giant neural networks using pipeline parallelism. In: Wallach H, Larochelle H, Beygelzimer A, et~al (eds) Advances in Neural Information Processing Systems, vol~32. Curran Associates, Inc., \doi{10.48550/arXiv.1811.06965}

\bibitem[{{Hunmin Yang, Se-Yoon Oh, Ki-Jung Ryu}(2019)}]{multigpu_accel_spark}
{Hunmin Yang, Se-Yoon Oh, Ki-Jung Ryu} (2019) {Accelerating distributed deep learning inference on multi-GPU with Hadoop Spark}. \urlprefix\url{https://developer.download.nvidia.com/video/gputechconf/gtc/2019/presentation/s9343-accelerating-distributed-deep-learning-inference-on-multi-gpu-with-hadoop-spark_V2.pdf}

\bibitem[{{Intel Corp.}(2017)}]{movidius_tech}
{Intel Corp.} (2017) {Intel Movidius Myriad X Vision Processing Unit (VPU) with Neural Compute Engine}. \urlprefix\url{https://www.intel.com/content/www/us/en/products/docs/processors/movidius-vpu/myriad-x-product-brief.html}

\bibitem[{Jahanshahi et~al(2020)Jahanshahi, Sabzi, Lau, and Wong}]{multigpu_energy}
Jahanshahi A, Sabzi HZ, Lau C, et~al (2020) Gpu-nest: Characterizing energy efficiency of multi-gpu inference servers. IEEE Computer Architecture Letters 19(2):139--142. \doi{10.1109/LCA.2020.3023723}

\bibitem[{James et~al(2019)James, Sirakoulis, and Roy}]{James2019}
James A, Sirakoulis GC, Roy K (2019) {Smart cameras everywhere: AI vision on edge with emerging memories}. In: IEEE Int. Conf. on Electronics, Circuits and Systems (ICECS), pp 422--425, \doi{10.1109/ICECS46596.2019.8965029}

\bibitem[{Jouppi et~al(2017)Jouppi, Young, Patil, and Patterson}]{Jouppi2017}
Jouppi NP, Young C, Patil N, et~al (2017) {In-Datacenter Performance Analysis of a Tensor Processing Unit}. In: Int. Sym. on Computer Architecture, p 1–12, \doi{10.1145/3079856.3080246}

\bibitem[{Jouppi et~al(2020)Jouppi, Yoon, Kurian, Li, Patil, Laudon, Young, and Patterson}]{DSA_supercomp}
Jouppi NP, Yoon DH, Kurian G, et~al (2020) A domain-specific supercomputer for training deep neural networks. Commun ACM 63(7):67–78. \doi{10.1145/3360307}

\bibitem[{Kang and Somtham(2022)}]{Kang2022}
Kang P, Somtham A (2022) An evaluation of modern accelerator-based edge devices for object detection applications. Mathematics 10(22). \doi{10.3390/math10224299}

\bibitem[{Kim et~al(2023)Kim, Lee, Kim, Park, Yoo, Kwon, and Lee}]{Kim2023}
Kim J, Lee JH, Kim S, et~al (2023) Memory-efficient fine-tuning of compressed large language models via sub-4-bit integer quantization. In: Oh A, Naumann T, Globerson A, et~al (eds) Advances in Neural Information Processing Systems, vol~36. Curran Associates, Inc., pp 36187--36207, \urlprefix\url{https://proceedings.neurips.cc/paper_files/paper/2023/file/7183f4fc87598f6c6e947b96714acbd6-Paper-Conference.pdf}

\bibitem[{Kumar et~al(2021)Kumar, Bradbury, Young, Wang, Levskaya, Hechtman, Chen, Lee, Deveci, Kumar, Kanwar, Wang, Wanderman-Milne, Lacy, Wang, Oguntebi, Zu, Xu, and Swing}]{kumar2021exploring}
Kumar S, Bradbury J, Young C, et~al (2021) Exploring the limits of concurrency in ml training on google tpus. \doi{10.48550/arXiv.2011.03641}

\bibitem[{Li et~al(2020)Li, Zeng, Zhou, and Chen}]{Li2019}
Li E, Zeng L, Zhou Z, et~al (2020) {Edge AI: On-Demand Accelerating Deep Neural Network Inference via Edge Computing}. IEEE Trans on Wireless Communications 19(1):447--457. \doi{10.1109/TWC.2019.2946140}

\bibitem[{Li et~al(2021)Li, Liang, Li, Xu, Jia, and Guo}]{Li2021}
Li J, Liang W, Li Y, et~al (2021) Throughput maximization of delay-aware dnn inference in edge computing by exploring dnn model partitioning and inference parallelism. IEEE Transactions on Mobile Computing 22(5):3017--3030. \doi{10.1109/TMC.2021.3125949}

\bibitem[{Li et~al(2023)Li, Zheng, Zhong, Liu, Sheng, Jin, Huang, Chen, Zhang, Gonzalez, and Stoica}]{Zhuohan2023}
Li Z, Zheng L, Zhong Y, et~al (2023) {AlpaServe}: Statistical multiplexing with model parallelism for deep learning serving. In: 17th USENIX Symposium on Operating Systems Design and Implementation (OSDI 23). USENIX Association, Boston, MA, pp 663--679, \urlprefix\url{https://www.usenix.org/conference/osdi23/presentation/li-zhouhan}

\bibitem[{Libutti et~al(2020)Libutti, Igual, Pi{\~n}uel, Giusti, and Naiouf}]{Libutti2020}
Libutti L, Igual FD, Pi{\~n}uel L, et~al (2020) {Benchmarking Performance and Power of USB Accelerators for Inference with MLPerf}. In: Workshop Accelerated Mach. Learn, p 1–15

\bibitem[{Ma et~al(2023)Ma, Fang, and Wang}]{Ma2013}
Ma X, Fang G, Wang X (2023) Llm-pruner: On the structural pruning of large language models. In: Oh A, Naumann T, Globerson A, et~al (eds) Advances in Neural Information Processing Systems, vol~36. Curran Associates, Inc., pp 21702--21720, \urlprefix\url{https://proceedings.neurips.cc/paper_files/paper/2023/file/44956951349095f74492a5471128a7e0-Paper-Conference.pdf}

\bibitem[{McEnroe et~al(2022)McEnroe, Wang, and Liyanage}]{McEnroe2022}
McEnroe P, Wang S, Liyanage M (2022) {A Survey on the Convergence of Edge Computing and AI for UAVs: Opportunities and Challenges}. IEEE Internet of Things Journal 9. \doi{10.1109/JIOT.2022.3176400}

\bibitem[{Mohammed et~al(2020)Mohammed, Joe-Wong, Babbar, and Di~Francesco}]{Mohammed2020}
Mohammed T, Joe-Wong C, Babbar R, et~al (2020) Distributed inference acceleration with adaptive dnn partitioning and offloading. In: IEEE INFOCOM 2020-IEEE Conference on Computer Communications, IEEE, pp 854--863, \doi{10.1109/INFOCOM41043.2020.9155237}

\bibitem[{Murshed et~al(2021)Murshed, Murphy, Hou, Khan, Ananthanarayanan, and Hussain}]{Murshed2022}
Murshed MGS, Murphy C, Hou D, et~al (2021) {Machine Learning at the Network Edge: A Survey}. ACM Comput Surv 54(8). \doi{10.1145/3469029}

\bibitem[{Narayanan et~al(2019)Narayanan, Harlap, Phanishayee, Seshadri, Devanur, Ganger, Gibbons, and Zaharia}]{Narayanan2019}
Narayanan D, Harlap A, Phanishayee A, et~al (2019) Pipedream: generalized pipeline parallelism for dnn training. In: Proceedings of the 27th ACM symposium on operating systems principles, pp 1--15, \doi{10.1145/3341301.3359646}

\bibitem[{Nikolić et~al(2022)Nikolić, Dimitrijević, Nikolić, and Stojcev}]{Nikolic2022}
Nikolić GS, Dimitrijević BR, Nikolić TR, et~al (2022) {A Survey of Three Types of Processing Units: CPU, GPU and TPU}. In: Int. Scientific Conf. on Information, Communication and Energy Systems and Technologies (ICEST), pp 1--6, \doi{10.1109/ICEST55168.2022.9828625}

\bibitem[{Parashar et~al(2020)Parashar, Abraham, Chaudhary, and Rajendiran}]{Parashar2020}
Parashar A, Abraham A, Chaudhary D, et~al (2020) Processor pipelining method for efficient deep neural network inference on embedded devices. In: 2020 IEEE 27th International Conference on High Performance Computing, Data, and Analytics (HiPC), IEEE, pp 82--90, \doi{10.1109/HiPC50609.2020.00022}

\bibitem[{Raj and Sekhar(2020)}]{Raj2020}
Raj P, Sekhar C (2020) {Comparative Study on CPU, GPU and TPU}. Int Journal of Computer Science and Information Technology for Education 5:31--38. \doi{10.21742/IJCSITE.2020.5.1.04}

\bibitem[{Ren et~al(2022)Ren, Qu, Dong, Jing, Sun, Wu, and Guo}]{Ren2022}
Ren W, Qu Y, Dong C, et~al (2022) {A Survey on Collaborative DNN Inference for Edge Intelligence}. \doi{10.48550/ARXIV.2207.07812}, \urlprefix\url{https://arxiv.org/abs/2207.07812}

\bibitem[{Renda et~al(2020)Renda, Frankle, and Carbin}]{Renda2020}
Renda A, Frankle J, Carbin M (2020) Comparing rewinding and fine-tuning in neural network pruning. In: International Conference on Learning Representations

\bibitem[{Sedgewick and Wayne(2011)}]{sedgewick2011}
Sedgewick R, Wayne KD (2011) Algorithms, 4th edn., Addison-Wesley Professional, pp 661--666

\bibitem[{Seshadri et~al(2021)Seshadri, Akin, Laudon, Narayanaswami, and Yazdanbakhsh}]{Seshadri2021}
Seshadri K, Akin B, Laudon J, et~al (2021) {An Evaluation of Edge TPU Accelerators for Convolutional Neural Networks}. \doi{10.48550/ARXIV.2102.10423}, \urlprefix\url{https://arxiv.org/abs/2102.10423}

\bibitem[{Sun and Kist(2021)}]{Sun2021}
Sun Y, Kist AM (2021) {Deep Learning on Edge TPUs}. \doi{10.48550/ARXIV.2108.13732}, \urlprefix\url{https://arxiv.org/abs/2108.13732}

\bibitem[{Thalluri et~al(2021)Thalluri, Venkat, Prasad, Kumar, Kumar, Sarma, and Adapa}]{Thalluri2021}
Thalluri LN, Venkat SN, Prasad CVVD, et~al (2021) {Artificial Intelligence Enabled Smart City IoT System using Edge Computing}. In: Int. Conf. on Smart Electronics and Communication (ICOSEC), pp 12--20, \doi{10.1109/ICOSEC51865.2021.9591732}

\bibitem[{Varghese et~al(2021)Varghese, Wang, Bermbach, Hong, Lara, Shi, and Stewart}]{Varghese2021}
Varghese B, Wang N, Bermbach D, et~al (2021) {A Survey on Edge Performance Benchmarking}. ACM Comput Surv 54(3). \doi{10.1145/3444692}

\bibitem[{Villarrubia et~al(2023)Villarrubia, Costero, Igual, and Olcoz}]{Villarrubia2023}
Villarrubia J, Costero L, Igual FD, et~al (2023) Improving inference time in multi-{TPU} systems with profiled model segmentation. In: 2023 31th Euromicro International Conference on Parallel, Distributed and Network-Based Processing (PDP), pp 84--91, \doi{10.1109/PDP59025.2023.00020}

\bibitem[{Wu et~al(2023)Wu, Gao, Yu, Zhou, Yang, and Wang}]{Wu2023}
Wu L, Gao G, Yu J, et~al (2023) Pdd: partitioning dag-topology dnns for streaming tasks. IEEE Internet of Things Journal \doi{10.1109/JIOT.2023.3323520}

\bibitem[{Xiang and Kim(2019)}]{Xiang2019}
Xiang Y, Kim H (2019) Pipelined data-parallel cpu/gpu scheduling for multi-dnn real-time inference. 2019 IEEE Real-Time Systems Symposium (RTSS) pp 392--405. \doi{10.1109/RTSS46320.2019.00042}

\bibitem[{Zeng et~al(2021)Zeng, Chen, Zhou, Yang, and Zhang}]{Zeng2021}
Zeng L, Chen X, Zhou Z, et~al (2021) Coedge: Cooperative dnn inference with adaptive workload partitioning over heterogeneous edge devices. IEEE/ACM Transactions on Networking 29(2):595--608. \doi{10.1109/TNET.2020.3042320}

\bibitem[{Zhou et~al(2023)Zhou, Li, Wang, Min, and Wu}]{Zhou2023}
Zhou H, Li M, Wang N, et~al (2023) Accelerating deep learning inference via model parallelism and partial computation offloading. IEEE Transactions on Parallel and Distributed Systems 34(2):475--488. \doi{10.1109/TPDS.2022.3222509}

\bibitem[{Zhou et~al(2019)Zhou, Wang, Ota, and Dong}]{Zhou2019}
Zhou J, Wang Y, Ota K, et~al (2019) Aaiot: Accelerating artificial intelligence in iot systems. IEEE Wireless Communications Letters 8(3):825--828. \doi{10.1109/LWC.2019.2894703}

\bibitem[{Zhou et~al(2018)Zhou, Moosavi-Dezfooli, Cheung, and Frossard}]{Zhou2018}
Zhou Y, Moosavi-Dezfooli SM, Cheung NM, et~al (2018) Adaptive quantization for deep neural network. Proceedings of the AAAI Conference on Artificial Intelligence 32(1). \doi{10.1609/aaai.v32i1.11623}

\end{thebibliography}
